\documentclass[10pt, journal, letterpaper]{IEEEtran}

\usepackage{cite}

\usepackage{multirow} 
\usepackage{algpseudocode}
\usepackage{algorithm}
\usepackage{rotating}
\usepackage{kantlipsum} 
\usepackage{commath}
\allowdisplaybreaks
\usepackage{mathtools}  
\usepackage{verbatim}
\usepackage{xr-hyper} 
\usepackage{enumitem}
\usepackage{multirow}
\usepackage[table,xcdraw]{xcolor}
\usepackage{array,arydshln}
\usepackage{graphicx,booktabs}
\usepackage{longtable}
\usepackage{balance}
\usepackage{flushend}
\usepackage{epstopdf}
\usepackage{wrapfig}
\usepackage{latexsym}
\usepackage{amssymb}
\usepackage{amsthm}
\usepackage{amsfonts}
\usepackage{amsmath} 
\usepackage{graphicx}
\usepackage{latexsym}
\usepackage{booktabs}
\usepackage{support-caption} 
\usepackage[style=base]{caption}
\usepackage{subcaption} 
\usepackage{breqn}

\algnewcommand\algorithmicinput{\textbf{INPUT:}}
\algnewcommand\INPUT{\item[\algorithmicinput]}
\algnewcommand\algorithmicoutput{\textbf{OUTPUT:}}
\algnewcommand\OUTPUT{\item[\algorithmicoutput]}
\usepackage[table]{xcolor}
\usepackage{graphicx}
\usepackage{mathtools}
\usepackage{enumitem,kantlipsum}
\usepackage{adjustbox}
\usepackage[thinlines]{easytable}

\algdef{SE}[DOWHILE]{Do}{doWhile}{\algorithmicdo}[1]{\algorithmicwhile\ #1}
\makeatletter
\algnewcommand{\LineComment}[1]{\Statex \hskip\ALG@thistlm \texttt{#1}}
\makeatother

\newlength\mylength
\usepackage{color}
\colorlet{BLUE}{blue}
\usepackage{colortbl}
\definecolor{LightCyan}{RGB}{155, 227, 247}

\newcommand{\dq}[1]{``#1''}

\newif\ifcommentson
\commentsontrue

\newif\ifextended
\newif\ifshortver

\extendedtrue

\newcommand{\extended}[1]{\ifextended \ifshortver \textcolor{purple}{#1} \else \textcolor{black}{#1} \fi  \fi}
\newcommand{\shortver}[1]{\ifshortver \ifextended \textcolor{blue}{#1} \else \textcolor{black}{#1} \fi \fi}

\newcommand{\optional}[1]{\ignorespaces}

\newif\ifrevisionactive
\newif\ifshowdeleted
\revisionactivetrue

\begin{document}
\title{Optimal Estimation of Link Delays based on End-to-End Active Measurements} 

\author{M.M Tajiki, S.H Ghasemi Petroudi, Stefano Salsano, Steve Uhlig,
Ignacio Castro
\IEEEcompsocitemizethanks{\protect
\IEEEcompsocthanksitem Mohammad Mahdi Tajiki, Steve Uhlig, and Ignacio Castro are with School of electronic engineering and computer science Queen Mary, University of London, UK - E-mail: \{m.tajiki, steve.uhlig, i.castro\}@qmul.ac.uk\protect
\IEEEcompsocthanksitem Stefano Salsano is with the Department of Electronic Engineering at the University of Rome – Tor Vergata and the Consorzio Nazionale Interuniversitario per le Telecomunicazioni (CNIT) – Rome, Italy E-mail: stefano.salsano@uniroma2.it\protect
\IEEEcompsocthanksitem  Seyed Hesamedin Ghasemi Petroudi is with the ECE Department, Tarbiat Modares University, Tehran, Iran - Email: h.qasemi@modares.ac.ir
}
\vspace{2ex}
\textbf{\\Extended version of submitted paper - Version 1 - December 2020}
\vspace{-4ex}
}

\date{December 2020}
\maketitle	
\begin{abstract}
Current IP based networks support a wide range of delay-sensitive applications such as live video streaming of network gaming. Providing an adequate quality of experience to these applications is of paramount importance for a network provider. The offered services are often regulated by tight Service Level Agreements (SLAs) that needs to be continuously monitored.
Since the first step to guarantee a metric is to measure it, delay measurement becomes a fundamental operation for a network provider. In many cases, the operator needs to measure the delay on all network links. We refer to the collection of all link delays as the Link Delay Vector (LDV). Typical solutions to collect the LDV impose a substantial overhead on the network. In this paper, we propose a solution to measure the LDV in real-time with a low-overhead approach. In particular, we inject some flows into the network and infer the LDV based on the delay of those flows. To this end, the monitoring flows and their paths should be selected minimizing the network monitoring overhead. In this respect, the challenging issue is to select a proper combination of flows such that by knowing their delay it is possible to solve a set of a linear equation and obtain a unique LDV. This combination of monitoring flows should be optimal according to some criteria and should meet some feasibility constraints. We first propose a mathematical formulation to select the optimal combination of flows, in form of Integer Linear Programming (ILP) problem. Then we develop a heuristic algorithm to overcome the high computational complexity of existing ILP solvers. As a further step, we propose a meta-heuristic algorithm to solve the above-mentioned equations and infer the LDV. The challenging part of this step is the volatility of link delays. The proposed solution is evaluated over real-world emulated network topologies using the Mininet network emulator. Emulation results show the accuracy of the proposed solution with a negligible networking overhead in a real-time manner.
\end{abstract}	

\section{Introduction}
Due to the increasing demand for data communication over computer networks, network administrators have to continually improve the way they manage their resources. One of the most promising technologies that offers many advantages for network management and monitoring is Software-Defined Networking (SDN). SDN provides a powerful solution to manage and monitor network traffic, based on the separation of control plane and data plane. More precisely, since the controller can get customized statistics from SDN nodes it is possible for the controller to view the status of the network from a centralized perspective. 
Although SDN brings a lot of benefits, still there are challenging parts to deal with. Meeting QoS constraints for delay sensitive applications is among the most important issue that should be addressed. QoS is especially important in networks where the capacity is a limited resource, for example in cellular data communication.

The above-mentioned issue attracts more attention in modern networks such as 5G network where there are many delay-sensitive applications. In other words, reducing network communication delay plays an important role in improving service quality. This is because there are many delay-sensitive services such as video streaming, VNF re-placement, multiplayer gaming, factory robots, and self-driving cars. For many of these high bit-rate delay-sensitive traffics, it is critical to have good network performance in order to provide QoS and guarantee an adequate QoE for the end-user. Taking this into account, delay management receives a great interest in modern computer networks. The first step to manage the delay is to measure it. This can be done using different approaches based on the main purpose of the measurement. One of the most valuable types of delay measurement is the measurement of the link delay vector (LDV). The Link Delay vector is the vector of the unidirectional delay between every pair of adjacent nodes (i.e. the vector of the delay on each unidirectional link). 

By \dq{link delay} we actually mean the delay from the moment a packet has been fully received by a node to the instant in which it has been fully transmitted on the outgoing link. Under this hypothesis, the link delay can be split in four parts: processing delay, queuing delay, transmission delay, propagation delay. The propagation delay is fixed, its constant value depends on the physical media and on the link length. The transmission delay depends on the line rate and on the packet size. Assuming a given packet size, both the transmission and propagation delay can be considered as fixed and are common between packets (of the same size!) passing through a route. On the other hand, processing delays and queuing delays are the variable parts of each link delay. In particular the queuing delay, which depends on the link utilization, is the most important source of the overall delay variability.

Considering that the link delay is variable with time, it is useful to measure it continuously, so that the network operator can promptly react when the delay on some links is higher than expected. In this paper, we use the notation of \textit{Real-Time} LDV (RT-LDV) to highlight the need of dynamically and continuously monitor the vector of link delays in a network.

Several applications can benefit from the knowledge of LDV or RT-LDV, among them we find: SLA-aware traffic engineering, Network troubleshooting and diagnostic. Regarding SLA-aware traffic engineering, the link delay vector is an invaluable input to do QoS-aware traffic engineering. Thanks to the centralized resource administration of SDN networks, flows can be routed via low delay links using these metrics. There are lots of works using network statistics to provide SLA-based routes such as \cite{kamoun2018ip,tajiki2017optimal,lin2018dte}. Another application of delay vector is in network troubleshooting and diagnostic. Due to the distributed nature of networks, troubleshooting is extremely hard. Many failures are considered as network \dq{problems} while they are not. Researches show that about 50$\%$ of these \dq{network} problems are not caused by the network~\cite{guo2015pingmesh}. However it is not easy to tell if a \dq{network} problem is indeed caused by network failures or not. For example some components can only be reached intermittently, trying to use them may be mistakenly considered as a link/router failure. Another field that can take the benefits of knowledge on delay vector is Multi-objective Resource Allocation. In some resource management scenarios, the network controller should do a cross layer resource allocation, e.g., in \cite{tajiki2018energy} the problems of traffic engineering and VNF placement are considered as a unique problem to minimize the network energy consumption while the flows receive services based on SLA agreement. This means that the controller can exploit the aforementioned link delay vector to provide a better service and simultaneously minimize the energy consumption.

There are several questions on how to measure the delay vector, e.g., how to handle multiple links (link bundles) along two networking nodes? How to handle multiple paths between two nodes (e.g, if using ECMP)? What is the possibility and the cost to change end-hosts (i.e. in case of using Host-based solutions to perform LDV measurements)? Shall hosts be engaged in the process of network monitoring (\cite{guo2015pingmesh})? or is it better to keep the hosts untouched and do the measurements inside the network? Throughout the remaining sections, we answer the above-mentioned questions. Our main goals in this paper can be expressed as follows: 
\begin{itemize}
\item Measuring the Link Delay Vector (LDV). 
\item Keeping the monitoring overhead below a predefined threshold. 
\item Minimizing the involvement of end hosts (servers/clients) in the measurements.
\item Being dynamic and providing real-time monitoring. 
\item Developing an efficient implementation of the proposed solution and offering it as a tool for other researchers.
\end{itemize}
It should be mentioned that there are many challenges around LDV measurement problem, however, we believe these goals are the high priority ones.
To meet the above goals, we propose a novel Active Network Monitoring Architecture called Efficient Link Delay Measurement (ELDM). The proposed architecture, provides real-time delay measurements for SDN-based networks. We consider that the routing devices are SDN-enabled so that we can impose the routing of the monitoring flows (flows injected to the network to perform LDV measurement). In other words, we facilitate the process of active monitoring in SDN through this architecture. The same way, SDN facilitates the process of network management by its architecture. In our approach, the end-to-end delay of a set of monitoring flows is measured and based on that information, per-link delay (which is the sum of processing, transmission, queuing and propagation delays) is inferred. More precisely, we mathematically formulate the problem of choosing a set of monitoring flows to be injected to the network so that the link delay vector can be inferred using the delays measured on the injected flows. The formulation is in form of Integer Linear Programming (ILP) and could be solved with existing tools like those listed in~\cite{wikipedia_2020_listOptimization_software}. However, in order to cope with the high computational complexity of optimizing an ILP problem, a heuristic algorithm is proposed to solve the problem in a real-time manner. Thereafter, several flows are injected to the network and their end-to-end delays are measured. The number of flows and the selected path for each of them should be chosen in a way that by solving a system of linear equations the delays of desired links could be specified. It is important to prevent the monitoring flows from using more than a predefined fraction of the link capacity. To have a solvable multi-variables multi-equations problem, none of the flows delays should be inferable from the delay of one or more of other flows\footnote{To have a unique solution when there are n-unknowns, we need n-equations that are not inferable from each other.}. Besides, the path length (i.e. hop count) of each monitoring flow should be less than a predefined value to keep the measurement accurate\footnote{Different probes are traversing links in different time stamps and the links' delays are fluctuating. Therefore, increasing the length of paths may cause a tiny inconsistency in the measured values. As we are using meta-heuristic, this inconsistency will be tolerated, however, keeping the length of the paths low will decrease the chance of these types of inconsistency.}.
In brief, the contributions of this paper are as follows:
\begin{itemize}
    \item An active network monitoring architecture is proposed. The architecture provides a real-time delay measurement for SDN-based networks;
    \item  We measure per-link delay (which includes propagation and queuing delays);
    \item We mathematically formulate the joint problems of node placement and flow selection for active network monitoring. To this end, three metrics are optimized: length of monitoring flows, monitoring overhead, and accuracy;
    \item The non linear constraints appearing in the aforementioned optimization problem are linearized, therefore, obtaining Integer Linear Programming (ILP) model that can be solved with existing toolbars (like CPLEX \cite{cplex2009v12}) for small-sized network scenarios;
    \item In order to cope with the computational complexity and grant intelligence to the solution, the problem is broken down into two sub-problems. Then, heuristic and meta-heuristic approaches are exploited to solve the aforementioned sub-problems;
    \item We have open sourced all implementations and the codes are available at~\cite{monitoringCodeOurImplementation};
    \item Finally, the performance of the proposed solution is evaluated by measuring accuracy, execution time, and configuration overhead.
\end{itemize}

\section{Related Works}
\begin{figure}
    \centering
    \includegraphics[width=\columnwidth]{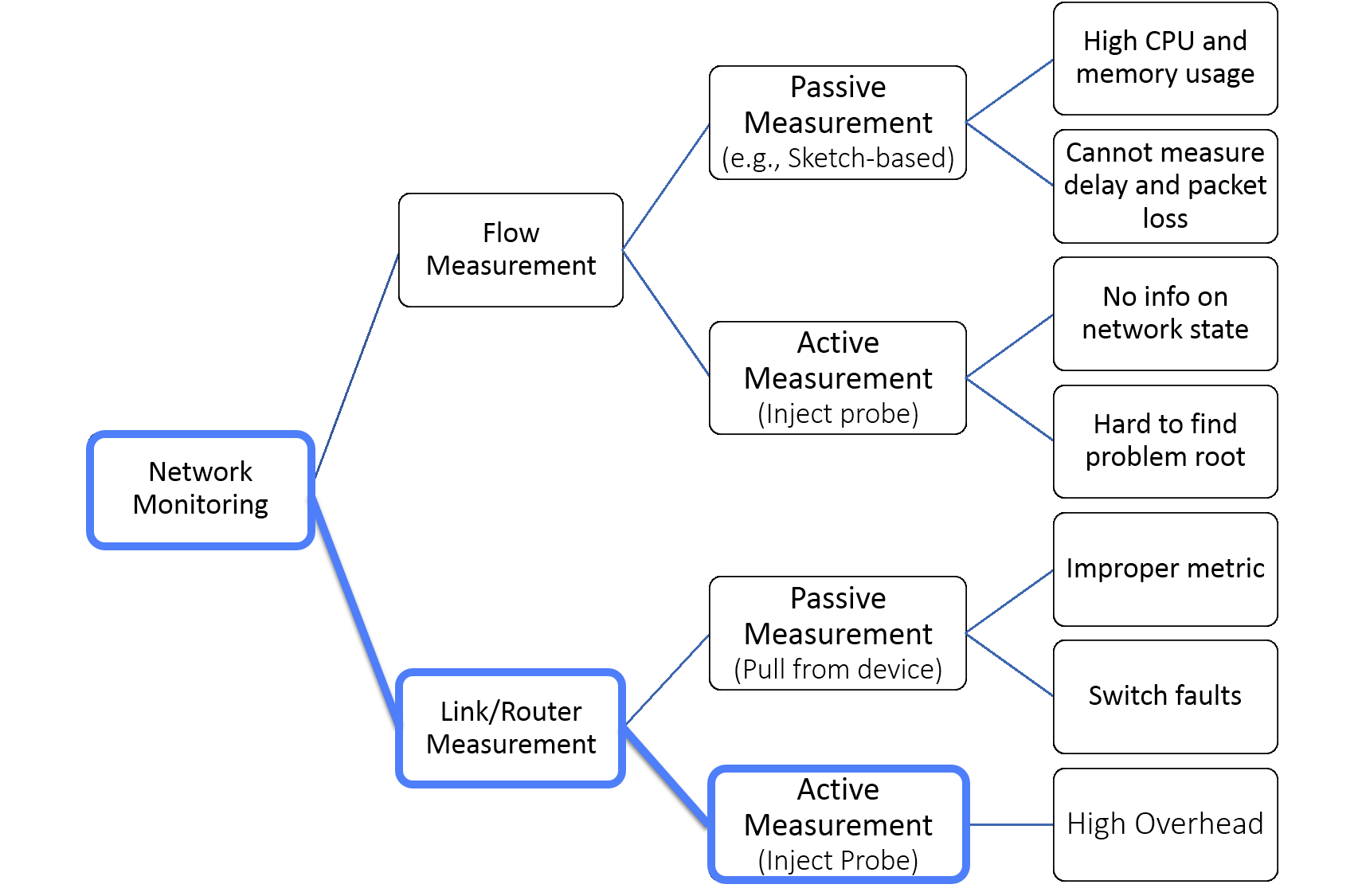}
    \caption{Different Monitoring Approaches.}
    \label{fig:differentMonitoring}
\end{figure}

Network monitoring can be divided into two main categories: I) those focusing on gathering flow level information (called Flow Measurement in Fig.~\ref{fig:differentMonitoring} \cite{roy2018cloud, guo2015pingmesh, gamage2019one,hernandez2007one,guingo2012distributed,bovy2002analysis,di2010new,van2014opennetmon}) and II) those focusing on network level information such as link delay vector or link load (called Link or Routing-Device Measurement in Fig.~\ref{fig:differentMonitoring} \cite{nakanishi2018route,naka2017route,gao2016accurate,tsang2003network,lawrence2006network,duffield2001network,gao2014domo}). Each of these categories has two sub-categories, passive and active methods. Passive measurement methods measure the desired metrics by observation, without injecting additional traffic in the form of probe packets. The main advantage of these methods is their zero network overhead (in terms of generating additional traffic). This means that passive measurement does not influence network performance. The main disadvantage of passive measurements is their reliance on installing in-network devices to monitor the traffic, which requires large investments and is not feasible for all networks. On the other hand, active measurements inject additional packets (referred as probe packets or monitoring flows) into the network, monitoring their behavior. Ping is a popular example for that in which ICMP packets are used to reliably compute a path’s round-trip time and determine end-to-end connection status.

There is another topic which is similar to our work in the goal, however, quite different in the output. Network tomography~\cite{vardi1996network} studies the internal characteristics of the network by measuring end-points data. Delay tomography aims at finding link delays using end-to-end probes sent from vantage points. In tomography, since there is no information about the internal network (e.g., topology and routing algorithm), this technique cannot accurately measure the link delays~\cite{lawrence2006networkNew}. In other words, delay tomography is quite different from our work because we know the topology thanks to the SDN based approach. Besides, we have access to the routing devices to enforce the desired routes to the monitoring flows.

Fig.~\ref{fig:differentMonitoring} represents the proposed categorization along with a brief indication to their challenging issues.
In this paper, the focus is on finding the delay vector of the known set of links, therefore, the state-of-the art works on active measurements are discussed and the papers related to the other categories are ignored.

\begin{figure*}
    \centering
    \includegraphics[width=\linewidth]{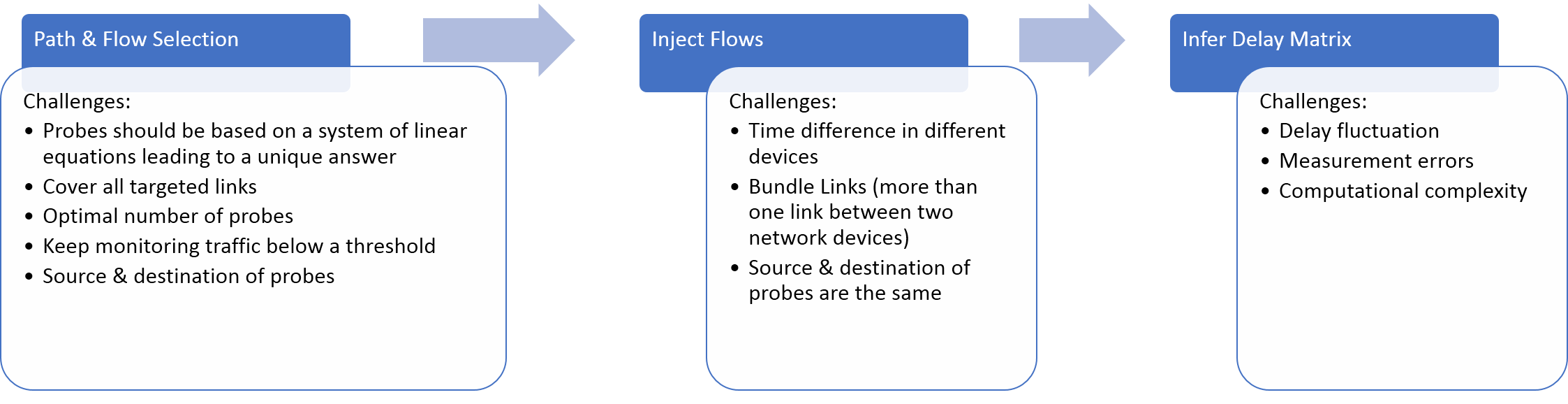}
    \caption{Steps taken in the proposed solution along with the challenges.}
    \label{fig:challenges}
\end{figure*}

Pingmesh~\cite{guo2015pingmesh} is a large-scale system for data centre network latency measurement and analysis. Pingmesh is an end-to-end measurement solution for delay monitoring. It leverages all the servers to launch HTTP or TCP pings to have comprehensive coverage on delay measurement. The algorithm forms several levels of complete graphs within a data centre. Similarly, servers within a rack collaborate in forming a complete graph. The top-of-rack (ToR) switches are considered as virtual nodes to let them form a second complete graph. Across data centres, Pingmesh establishes a third complete graph by considering each data centre as a virtual node. A central Pingmesh Controller undertakes the calculation of the complete graphs and related ping parameters. Although Pingmesh provides multiple monitoring wins at Microsoft (including detecting black holes and silent packet drops \cite{guo2015pingmesh}), an important question to ask is if this solution can be used for monitoring of virtual networks. The brief answer is no, motivating the authors of \cite{roy2018cloud} to develop VNET Pingmesh, an uplifted version of Pingmesh. This enhanced solution characterizes the precision and recalls to validate their effectiveness. However, they confess that they find it hard to do that without the ground truth gathered from customers. This is due both to the coverage issue they have with hybridized networks and because customers themselves may miss some issues.

There are several other works~\cite{zhang2020joint, zhang2019sensitive, yaseen2018synchronized} focusing on LDV measurement, however, their main drawback is the ultra high computational complexity of their solutions. The solution proposed in \cite{zhang2020joint} is joint consideration of both path delay monitoring and flow routing. In the first part, the authors focus on designing a monitoring mechanism to measure the link delay in heterogeneous networks. In this way, they exploit both link layer discovery protocol (LLDP) and Echo probing. Then, a dynamic routing algorithm is proposed to select optimized transmission paths based on the information of link latency. The RYU controller is deployed for feasibility and efficiency evaluation of the proposed solution. The authors of \cite{zhang2019sensitive} take an active measurement approach by establishing a buffer for the received packets. Their main focus is on end-to-end jitter measurement and they do not infer per link information from the gathered data. Yaseen et. al design a Synchronized Network Snapshot protocol in~\cite{yaseen2018synchronized}. Their primary goal is to collect a network-wide set of measurements. To have a meaningful measurement their design should be both causally consistent and approximately synchronous. 

The authors of \cite{gao2016accurate} focus on the problem of extracting per-hop delay for each packet from the end-to-end delay. They propose a per-link delay measurement by mathematically formulating the problem into a set of optimization sub-problems. Their solution adds an overhead header to every packet to construct the constraints of the optimization problems. By applying a relaxation and solving these optimization problems, their solution is able to estimate per-hop delays as well as giving an upper and lower bound for each link delay. This solution not only adds a lot of overhead but also has clock synchronization issue. Because of that the authors of \cite{nakanishi2018route} study synchronization-free delay measurement. In this way, they use theoretical analysis to cancel or minimize the error factors caused by clock asynchronism. However, removing or reducing this error ends up to a very high computational complexity which makes the algorithm hard to be applied for real-world networks. Besides, they have some considerations which rarely stands for real-world networks. It is worth mentioning that although we are seeking to solve the same problem \cite{gao2016accurate}~and~\cite{nakanishi2018route} are trying to, the solutions/approaches are completely different, i.e., our work is not an extension of these works. In our work, we use loops to overcome with the clock synchronization issue. Besides, we exploit SDN capabilities to propose a low overhead solution (no need to add extra headers).


\section{Active Delay Measurement}
In this section, we first depict an outline of the approach we have taken along with the challenges on the way. Fig.~\ref{fig:challenges} shows three main steps of the proposed solution and the corresponding issues of each step. 
The first step, called \textit{Path $\&$ Flow Selection}, is to define the flows, i.e., specifying the source, destination, and selected hops (links) throughout the network. This step is the most difficult one, including several challenges. The main challenge is that the assignment of resources (selecting paths for flows) should end up to a system of linear equations (or linear system) with a unique solution. A linear system is a collection of linear equations involving the same set of variables. The equations of a linear system should be independent meaning that none of the equations should be derived algebraically from the others. When the equations are independent, each equation contains new information about the variables, and removing any of the equations increases the size of the solution set. For linear equations, logical independence is the same as linear independence\shortver{(for more explanation refer to \cite{extendedversionofOptimalEstima2020})}. \extended{Fig.~\ref{fig:linear_systems_dependent} shows a sample where the equations are dependent in a 3D environment.
\begin{figure}[hbt!]
    \centering
    \begin{subfigure}{0.49\columnwidth}
          \centering
          \includegraphics[width=\columnwidth]{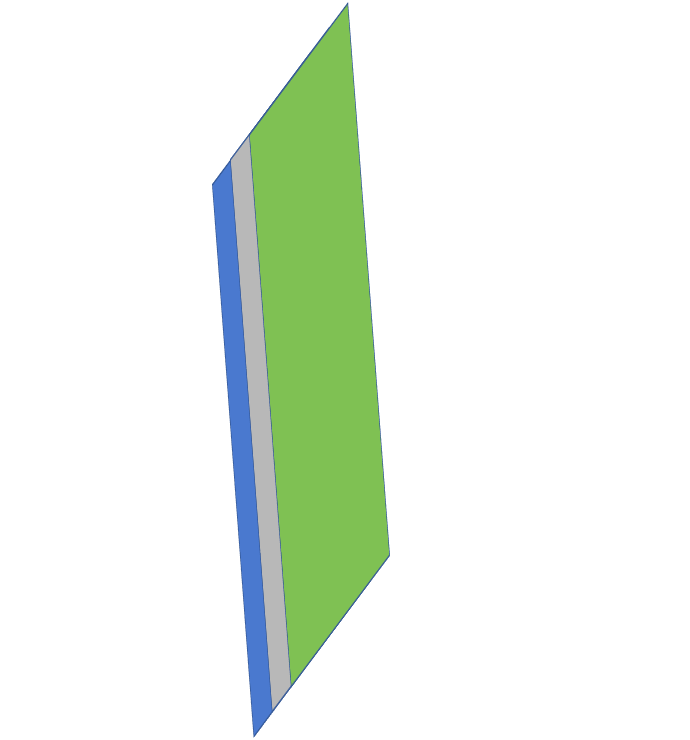}
          \caption{Dependent}
          \label{fig:linear_systems_dependent}
    \end{subfigure}
    \begin{subfigure}{0.49\columnwidth}
          \centering
          \includegraphics[width=\columnwidth]{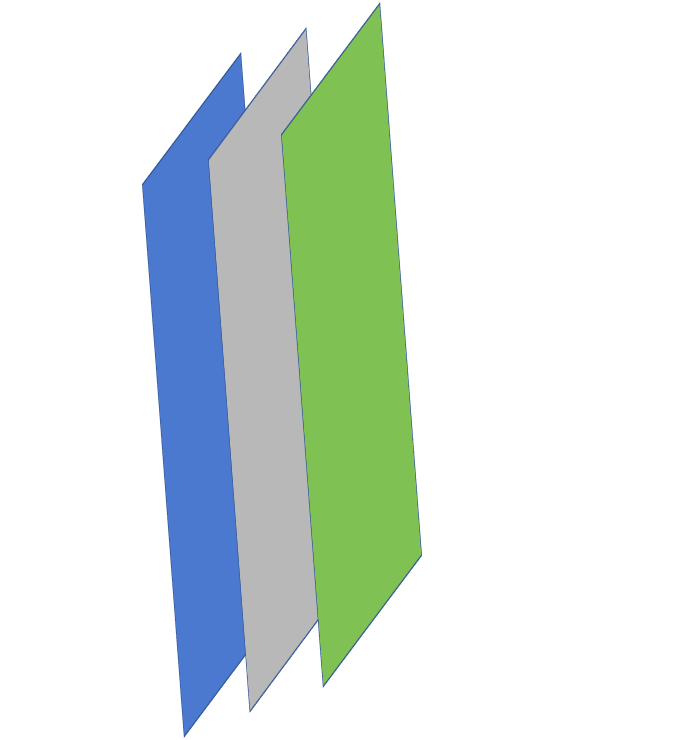}
          \caption{Inconsistent}
          \label{fig:linear_systems_inconsistent1}
    \end{subfigure}
    \begin{subfigure}{0.49\columnwidth}
          \centering
          \includegraphics[width=\columnwidth]{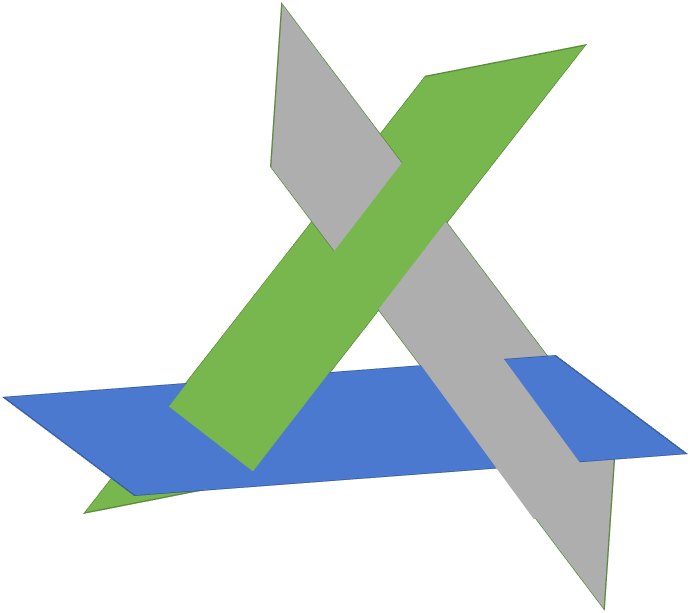}
          \caption{Inconsistent}
          \label{fig:linear_systems_inconsistent2}
    \end{subfigure}
    \begin{subfigure}{0.49\columnwidth}
          \centering
          \includegraphics[width=\columnwidth]{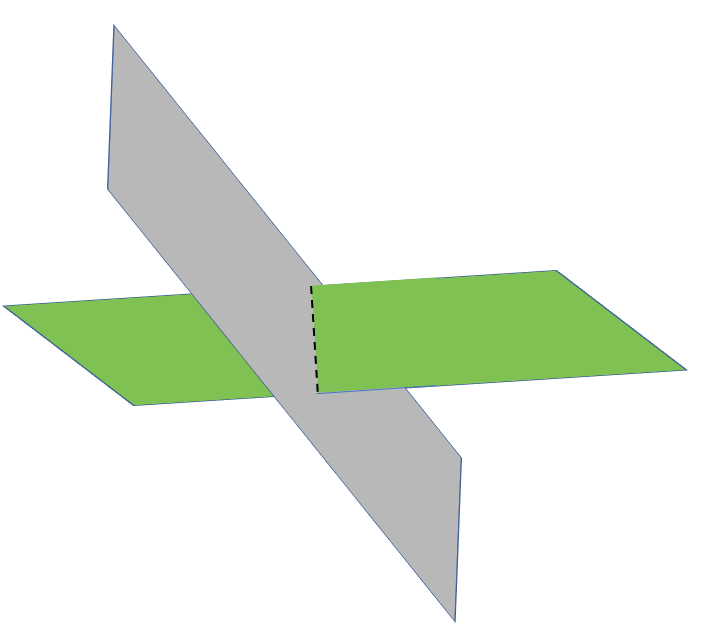}
          \caption{Many Answers}
          \label{fig:linear_systems_many_answers}
    \end{subfigure}
    \caption{Improper Linear Systems}
    \label{fig:linear_systems}
\end{figure}}
Another mandatory feature of the desired linear system is consistency. A linear system is consistent if it has a solution, and otherwise, it is said to be inconsistent. When the system is inconsistent, it is possible to derive a contradiction from the equations, that may always be rewritten as the statement 0 = 1. \extended{Fig.s~\ref{fig:linear_systems_inconsistent1}~and~\ref{fig:linear_systems_inconsistent2} are examples of inconsistency in linear equations.}

The desired linear system not only should be independent and consistent but also it should lead to one unique answer. \extended{For instance, the equations depicted in Fig.~\ref{fig:linear_systems_many_answers} are consistent and independent but leading to many possible answers. An example of a proper system of equations is illustrated in Fig.~\ref{fig:linear_system_perfect}, where solving the system leads to a unique answer.
\begin{figure}
    \centering
    \includegraphics[width=0.49\columnwidth]{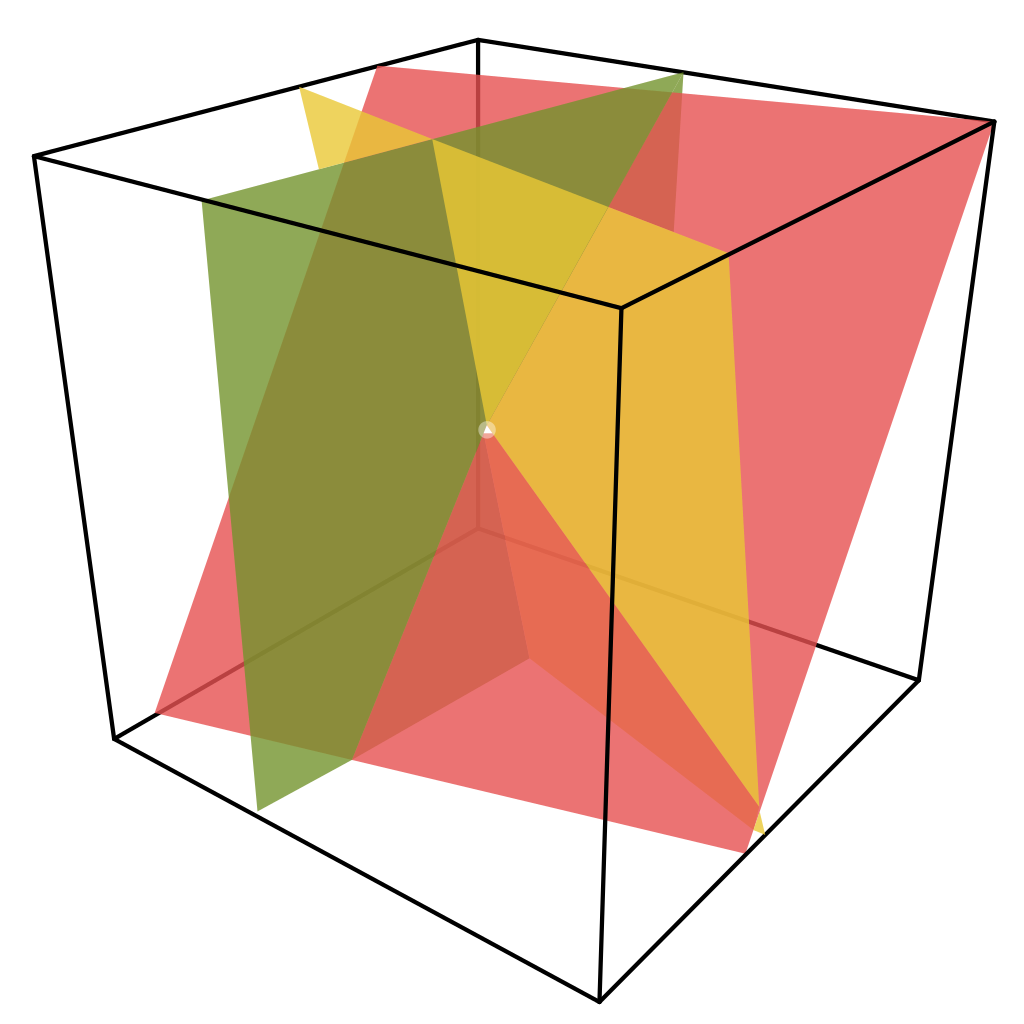}
    \caption{A proper system of linear equations with one unique answer.}
    \label{fig:linear_system_perfect}
\end{figure}}
Another challenge of \textit{path $\&$ flow selection} is to cover all targeted links meaning that all those links should appear in the linear system. In this way, increasing the length of the path each flow takes increases the measurement error while decreasing the length of the path increases the number of required flows. On the other hand, the number of flows should stay in a safe zone since decreasing the number of flows may result in a linear system with many possible answers while increasing the number of flows could end-up with too much monitoring traffic or an inconsistent linear system (due to measurement error). Therefore, the number of flows, their source~$\&$~destination, and the way they tour the network have a direct impact on the monitoring overhead.

The second step in the proposed solution is to inject the flows into the network in order to measure their end-to-end delay (called \textit{Inject flows} in Fig.~\ref{fig:challenges}). The main challenge of this part is to handle the time difference of monitoring devices. The error-tolerance of delay-measurement needs to be in a safe zone which is beyond the adjustment accuracy of computer devices. In another word, a computer clock can be thrown off by many factors, including network jitter, delays introduced by software, and even the environmental conditions in which the computer is operating. This means that network delay has an impact on the time accuracy in monitoring devices. Therefore, if the source and destination of a flow are two different entities then the one-way delay is not accurately computable due to the time difference of source and destination. It is worth mentioning that GPS based clock synchronization~\cite{tian2020high} could be a viable solution, however, we prefer to work on a solution that does not require GPS module.
All things considered, the applicable way of measuring one-way delay based on the analysis of flows' end-to-end delay is to set the source and destination to one node. This assumption brings up some implementation issues which is handled in our developed code (all code is open access and available from~\cite{monitoringCodeOurImplementation}). The main challenge of implementing this idea is that a router/switch needs to route a flow twice: once when the flow reaches the router/switch for the first time and once on the way back (to have a loop). Therefore, a flow with the same 5-tuples (source IP/port, destination IP/port,  and protocol in use) should be behaved differently when it pass through the networking device for the first and second times.

The last step in the proposed solution is to infer the matrix of delay using the information gathered in the previous step. The main challenge, here, is the link delay fluctuation and measurement errors which may cause an inconsistency in the linear system. Obviously, the computational complexity of this part is very important as this part is running in a periodic manner (but step one is a one-time task which could be done offline). In the following, the mathematical formulation and the meta-heuristic algorithm proposed for \textit{Path $\&$ flow Selection} are described in sec.s~\ref{subsec:mathematical_formulation}~and~\ref{subsec:heuristic_alg}. The meta-heuristic algorithm proposed for inferring the delay vector is discussed in sec.~\ref{subsec:pso}.

\subsection{Proposed Architecture}\label{subsec:Architecture}
\begin{figure}
    \centering
    \includegraphics[width=\columnwidth]{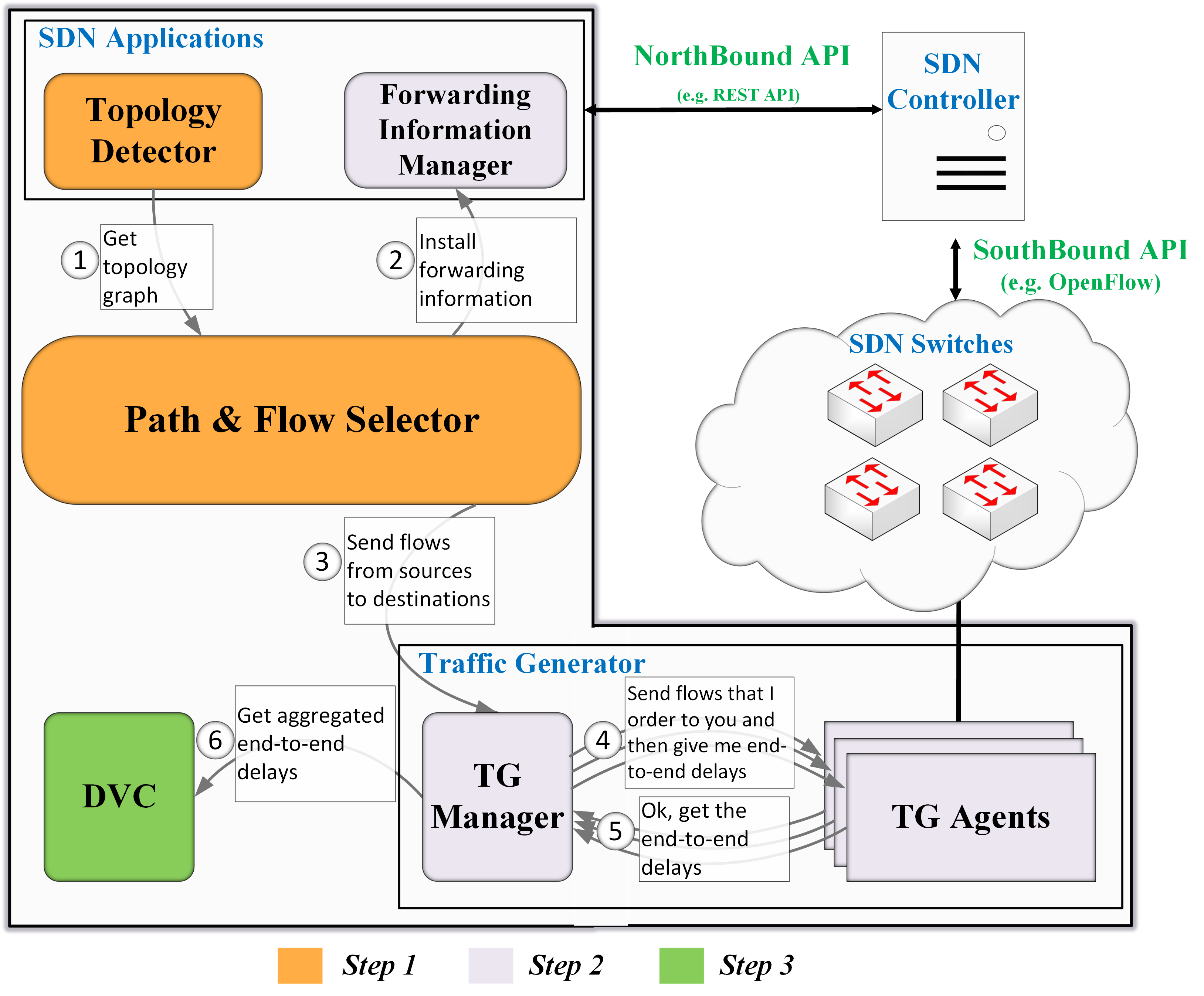}
    \caption{The proposed architecture.}
    \label{fig:Architecture}
\end{figure}

In this section, the proposed active monitoring architecture is discussed by illustrating the architectural components and providing an outline of the main considered components. We assume that a Network Operator is managing the network and it is able to allocate resources (e.g. processing capacity in nodes, bandwidth capacity on network links) to a set of monitoring flows that are intended to measure delays of links. The routing devices are SDN-enabled switches. We assume that there is a logically centralized SDN controller, connected to the set of SDN switches via a Southbound protocol. The main role of the SDN controller is to setup the forwarding tables in order to properly configure the packet forwarding. The SDN controller interacts with the switches and gathers information about the topology and on the network traffic. The SDN controller is connected to the controlling applications (SDN applications) through northbound APIs (which is REST API in Fig.~\ref{fig:Architecture}). As can be seen in Fig. \ref{fig:Architecture}, the proposed architecture has the following modules: 

\begin{itemize}
    \item \textit{Path $\&$ Flow Selector}: this module decides on the number of monitoring flows and specifies the required resources for each of these flows. In this context, the resources are links and the task is to perform route calculation for monitoring flows.
    \item \textit{Topology Discovery}: the main task of Topology Discovery is to periodically report the network topology to \textit{Path $\&$ Flow Selector}. When a new routing device enters the network, an existing node fails, or any changes take place in the network topology comes into the force, this module reports the changes to \textit{Path $\&$ Flow Selector}.
    \item \textit{Forwarding Information Manager (FIM)}: this module enforces networking equipment to act based on the decision made in the Application layer. To this end, \textit{FIM} exploits a northbound API to connect to the SDN controller. \textit{FIM} could be mapped in an SDN controller to increase the performance.
    \item \textit{Traffic Generators (TG Manager $\&$ TG Agent)}: these two modules generate the monitoring traffic based on the decision of \textit{Path $\&$ Flow Selector} module. Additionally, they measure the end-to-end delay for every monitoring flow. These two modules have nothing to do with network configuration. \textit{TG Manager} receives the list containing source/destination pair of every monitoring flows. Thereafter, it coordinates different instances of \textit{TG Agents}. On the other hand, \textit{TG Agent} which is running on monitoring nodes, generates the traffic based on the requested pair of source/destination. It is worth mentioning that there is only one instance of \textit{TG Manager} running while there are several instances of \textit{TG Agents} generating the monitoring traffic.
    \item \textit{Delay Vector Calculator (DVC)}: based on end-to-end delays and selected routes, this module infers delays of different links in the network. This should be done in a real-time manner and the algorithm should have a low computational complexity.
\end{itemize}

\subsection{Mathematical Formulation of Path $\&$ Flow Selection}\label{subsec:mathematical_formulation}
We consider an SDN-based network in which networking devices are programmed using a southbound protocol (e.g., reference~\cite{ventre2018sdn}). A few hosts originate the monitoring flows. 
For every monitoring flow, the controller should select a source, destination, and the set of links from the source to the destination. The length of the selected path for each flow should be less than a predefined threshold. All targeted links should be crossed by at least $m$ flows. The source and destination of a flow may be the same node. We assume a network with $N$ SDN-enabled switches. The network topology is represented with a matrix $B^{\max}_{N\times N}$, where the capacity of the link from the switch $i$ to the switch $j$ is stored in $b_{(i,j)}$. Similarly, $D_{N\times N}$ denotes the link delay, where $d_{(i,j)}$ stores the delay of the link from the switch $i$ to the switch $j$. The number of Monitoring flows in the network is denoted with $F$ and the maximum length of paths assigned is limited by $l$. Table~\ref{tab:notation} provides the symbols used though this paper along with a brief description.
\begin{table}[t]
	\caption{Main Notation.}\label{tab:notation}
	\centering
	\resizebox{\columnwidth}{!}{%
	\begin{tabular}{|c||l|l|} 
		\hline
		& \textbf{Symbol} & \textbf{Description}\\
		\hline\hline
		\multirow{13}{0.1cm}{\begin{sideways}\textbf{Parameters}\end{sideways}} 
		& $E$ & Number of links  \\\cline{2-3}
		& $N$ & Number of routing devices \\\cline{2-3}
		& $\mathcal{N}$ & Set of routing devices \\\cline{2-3}
        & $F$ & Number of monitoring flows \\\cline{2-3}
        & $\mathcal{F}$ & Set of monitoring flows \\\cline{2-3}
		& $B^{\max}_{(i,j)}$ & Matrix of link capacity\\ \cline{2-3}
	    & $D^{\max}_{f}$ & Maximum tolerable propagation \\ \cline{2-3}
	    & $q$ & Average rate of a monitoring flow  \\\cline{2-3}
		& $s^{f}$ & Vector of source switch\\\cline{2-3}
		& $d^{f}$ & Vector of destination switch\\\cline{2-3}
		& $x$ & Min number of monitoring flows on a link\\\cline{2-3}
		& $U$ & Sufficiently big number\\\cline{2-3}
		& $l$ & Max acceptable length for a monitoring flow\\\cline{2-3}
		& $A_{(i,j)}$ & Matrix of adjacency (link from switch i to j)\\ \hline\hline
		\multirow{7}{0.1cm}{\begin{sideways}\textbf{Variable}\end{sideways}}
		& $D_{(i,j)}$ & Links delay\\\cline{2-3}
		& ${\mu}$ & Monitoring traffic to total bandwidth ratio \\ \cline{2-3}
		& $R_{(i,j)}^f$ & Routing matrix \\\cline{2-3} 
		& $T^{(f,f')}$ & Intermediate variable (description in text) \\\cline{2-3} & $Y^{(f,f')}$ & Binary intermediate variable (description in text) \\\cline{2-3}
		& $P_{(i,j)}^f$ & Ordering matrix \\ \hline
    	\end{tabular}}
    \end{table}
\extended{In order to simplify the understanding of the notation, a sample of the proposed notation is presented in the following lines. A network topology with five routing devices can be represented with a $5\times5$ matrix as following:
\[B^{\max}=
  \begin{bmatrix}
    0 & b_{(1,2)} & b_{(1,3)} & 0 & 0\\
    b_{(2,1)} & 0 & 0 & b_{(2,4)} & 0\\
    b_{(3,1)} & 0 & 0 & 0 & b_{(3,5)}\\
    0 & b_{(4,2)} & 0 & 0 & b_{(4,5)}\\
    0 & 0 & b_{(5,3)} & b_{(5,4)} & 0
  \end{bmatrix}\]
\[D=
  \begin{bmatrix}
    \infty & d_{(1,2)} & d_{(1,3)} & \infty & \infty\\
    d_{(2,1)} & \infty & \infty & d_{(2,4)} & \infty\\
    d_{(3,1)} & \infty & \infty & \infty & d_{(3,5)}\\
    \infty & d_{(4,2)} & \infty & \infty & d_{(4,5)}\\
    \infty & \infty & d_{(5,3)} & d_{(5,4)} & \infty
  \end{bmatrix}\]}
  
The source and destination of flows are determined by the vectors $s^f$ and $d^f$ determine, respectively. In the proposed solution source and destination of a flow can be the same. Clearly this ends up to a loop in the routing path which is handled in our implementation (later on we will explain why loops may happen in this section). Except loops caused by equal source and destination, the other types of loops are not allowed (due to implementation limits). Matrix $A_{(i,j)}$ denotes whether a link from switch $i$ to switch $j$ exists or not. If there is a link then its value is 1 otherwise, 0. $x^t$: In each time slot, exactly $x^t$ monitoring flows travel through the link $(i,j)$. Matrices $R^f_{(i,j)}$ and $P^f_{(i,j)}$ denote the path selected for every monitoring flow. $R^f_{(i,j)}$, the main output of this formulation, shows the steps taken from source to destination without specifying the order\shortver{(for more explanation refer to \cite{extendedversionofOptimalEstima2020})}. \extended{As an example, let the source and destination of flow~1 be node one and node five respectively. 
\[R^1=
  \begin{bmatrix}
    0 & 1 & 0 & 0 & 0\\
    0 & 0 & 0 & 1 & 0\\
    0 & 0 & 0 & 0 & 0\\
    0 & 0 & 0 & 0 & 1\\
    0 & 0 & 0 & 0 & 0
  \end{bmatrix}\]
This means that the flow~1 starts from node one and directly goes to node two ($R^1_{(1,2)}=1$). Then the flows moves out to node four (since $R^1_{(2,4)}$ is one) and finally leave node four toward node five ($R^1_{(4,5)}=1$).} Similarly, $P^f_{(i,j)}$ is an intermediate variable showing the selected routing for every monitoring flow considering the ordering of links traversed. \extended{Therefore, matrix $P$ for the above mentioned flow is as following:
\[P^1=
  \begin{bmatrix}
    0 & 1 & 0 & 0 & 0\\
    0 & 0 & 0 & 2 & 0\\
    0 & 0 & 0 & 0 & 0\\
    0 & 0 & 0 & 0 & 3\\
    0 & 0 & 0 & 0 & 0
  \end{bmatrix}\]}
The mathematical formulation is provided in the following lines. The objective function is to minimize the monitoring overhead by reducing the maximum ratio of monitoring traffic to link capacity.
\begin{align}
    & \min{\mu}
\end{align}

Eq.~\ref{eq:use_existing_links} is used to ensure the links through a path do exists. In other words, flows are forced to use only existing links and do not jump from a router to another one if there is no direct link between them. To this end, the elements of routing matrix $R$ should always be less than or equal to corresponding elements in adjacency matrix $A$.
\begin{align}
    & R_{(i,j)}^f \leq A_{(i,j)}, \forall i, j \in \mathcal{N}, \forall f \in \mathcal{F} \label{eq:use_existing_links}
\end{align}

End-to-end delay may be undergone changes by different transient factors, therefore, to keep the measurement accuracy acceptable, it's better to use more than one flow to infer a link's delay. This means that having more than one flows traversing a link is desirable. Eq.~\ref{eq:flows_per_link} ensures that at least $x$ monitoring flows traverse every link. The appearance of adjacency matrix $A$ in the right side of Eq.~\ref{eq:flows_per_link} prevents the algorithm from using links that does not exists.
\begin{align}
    & \sum_{f=1}^{F}{R_{(i,j)}^{f}}\geq x*A_{(i,j)},   \forall i,j \in \mathcal{N} \label{eq:flows_per_link}
\end{align}

The net flow entering a node should be zero, except for the source, which "produces" flow, and the sink/destination, which "consumes" flow. This is usually referred to as "flow conservation constraints". In our mathematical formulation, the flow conservation concept is algebraically implemented in  Eq.~\ref{eq:flow_conservation}. However, here there is a small difference with common flow conservation constraints. In the proposed solution, it is possible to have a node as both source and destination of a flow. But due to challenges for routing in existence of loop, this case is usually unacceptable for most of routing algorithms, i.e., it is not considered in common flow conservation constraints. Therefore, the case where source and destination are the same is excepted in Eq.~\ref{eq:flow_conservation}. We then add a new constraint to handle this exception. Beside, practical solution to tackle the existence of loop for implementation step is discussed in the remainder of this paper.
\begin{align} \label{eq:flow_conservation}
    & \sum_{j=1}^N{\sum_{f=1}^F{R_{(i,j)}^{f}}}+\sum_{f=1}^F{\sum_{j=1}^N{R_{(j,i)}^{f}}}=\\\nonumber
    &~~~~~~~~~~~~~~~~~ \begin{cases}
                     1 & i=s^f~~~~ \And s^f\neq d^f\\
                    -1 & i=d^f~~~~ \And s^f\neq d^f\\
                     0 & i\neq s^f,d^f \And s^f\neq d^f\\
                    \end{cases},  \forall i \in \mathcal{N}
\end{align}

Eq.~\ref{eq:flow_conservation_replacement} substitutes the flow conservation constraint (Eq.~\ref{eq:flow_conservation}) when the source and destination of a flow are the same. In this case, the incoming and outgoing of every node should be zero, regardless of being a source, destination, middle node, or not traversed node. Eq.~\ref{eq:leavesource} ensures the flow leaves the source node.
\begin{align}
    & \sum_{j=1}^N{\sum_{f=1}^F{R_{(i,j)}^{f}}}+\sum_{f=1}^F{\sum_{j=1}^N{R_{(j, i)}^{f}}}= 0,~~~ s^f=d^f, \forall i \in \mathcal{N} \label{eq:flow_conservation_replacement} \\
    & \sum_{j=1}^N{\sum_{f=1}^F{R_{(i,j)}^{f}}} = \begin{cases}
                     1 & i=s^f\\
                     0 & i\neq s^f\\
                    \end{cases}~~~~~,\forall i \in \mathcal{N}\label{eq:leavesource}
\end{align}

One of the most important criteria that should be met is keeping the monitoring overhead as low as possible. There are several metrics to measure the overhead including the amount of monitoring traffic and the number of monitoring rules on routing devices. Eq.~\ref{eq:monitoring_overhead} ensures the monitoring flows use at most $\mu^t$ percent of link bandwidth.
\begin{align}\label{eq:monitoring_overhead}
    & \sum_{f=1}^F{R_{(i,j)}^{f}}\times q\leq \mu\times B_{(i,j)},   \forall i,j \in \mathcal{N}
\end{align}

There are two reasons for a loop to exists in a computer network: to meet a node twice or due to an error in the routing algorithm. If the loop is formed by an error in the routing algorithm then the path to a particular destination forms an infinite loop. This means that the nodes keep forwarding the packet to each other until an external reason (e.g., packet timer) intervene the routing. In our formulation, to prevent an unwanted loop creation, Eq.~\ref{eq:loop} is added. This equation prevents a loop in routing unless the source and destination are the same (this is a type of loop that does not make any problem).
\begin{align}\label{eq:loop}
    & \sum_{j=1}^N{R_{(i,j)}^{f}}\leq 1
\end{align}

Due to the dynamic nature of link delay, it is desired to have an upper bound on the number of links a monitoring flow traverse. To this end, Eq.~\ref{eq:flow_length} keeps the length of monitoring routes less than a predefined value.
\begin{align} \label{eq:flow_length}
    & \sum_{i=1}^N{\sum_{j=1}^N{R_{(i,j)}^{f}}}\leq l,     \forall f\in \mathcal{F}
\end{align}

Eq.~\eqref{eq:R_Zero} hints that if the value of matrix $A_{(i,j)}^f$ is zero, then $P^f_{(i,j)}$ becomes zero. It also indicates that the value of the ordering matrix should be higher or equal to the rerouting matrix.
\begin{align} \label{eq:R_Zero}
    & P_{(i,j)}^{f}\leq l\times R^f_{(i,j)},     \forall i,j \in \mathcal{N}, \forall f\in \mathcal{F}
\end{align}

The elements of ordering matrix specify the steps through the way from source to destination. Therefore, if the step number assigned to the flow entering a node $n$ is $X$ then step number assigned to that flow leaving node n should be $X+1$. This means that ordering matrix should always be equal to or greater than routing matrix. Briefly, the step number in ordering matrix should always increase by one after leaving a node. This criteria is guaranteed via Eq.~\ref{eq:step_P} in our formulation.
\begin{align} \label{eq:step_P}
    & \sum_{j=1}^N{P_{(i,j)}^{f}} = \sum_{j=1}^N{\left(P_{(j,i)}^f+R_{(j,i)}^f\right)}, \\\nonumber
    &~~~~~~~~~~~~(i\neq s^f ~or~ i\neq d^f), \forall i\in \mathcal{N}, \forall f\in \mathcal{F}
\end{align}

In our formulation, in order to deal with the time inconsistency between different nodes (in microsecond granularity), the source and destination of a flow may be the same. The following equations ensure that monitoring flows leave the source node and enter the destination node. This is because if the source and destination are the same the flow has reached the destination without leaving it. In other words, without the following equations, some flows will not leave the source.
\begin{align}
    & \sum_{j=1}^N{R^f_{(s^f,j)}}=1, \forall f\in \mathcal{F}\\
    & \sum_{i=1}^N{R^f_{(i,d^f)}}=1, \forall f\in \mathcal{F}
\end{align}

A solution for the proposed mathematical formulation may end up with similar routes assigned to flows with similar pairs of source $\&$ destination. Meaning that some flows are a duplication of others. Duplicating a flow (having two flows traversing similar nodes) do not add any extra information regarding the links' delay. However, this adds a new limit on the diversity of source $\&$ destination because completely different pairs of source $\&$ destination should be used for flows. Consequently, the number of required monitoring nodes to measure the delay of a set of links increases. To avoid this, the formulation should take diversity in routes as a criterion for flows with similar pairs of source $\&$ destination.
\begin{align}\label{eq:T}
    & T^{(f,f')} = \sum_{i=1}^N{\sum_{j=1}^N{\left(((N+1)\times i+j)\times R_{(i,j)}^f\right)}} - \nonumber\\ 
    &~~~~~\sum_{i=1}^N{\sum_{j=1}^N{\left(((N+1)\times i+j)\times R_{(i,j)}^{f'}\right)}}
\end{align}    
In order to simplify the formulation, intermediate variable $T^{(f,f')}$ is defined in Eq.~\ref{eq:T}. This variable compares the routes assigned to flows $f$ and $f'$ and computes a value for their similarity based on Eq.~\ref{eq:T}. 
\begin{align}
    & T^{(f,f')} \leq  Y^{(f,f')}\times U, \forall f,f'\in \mathcal{F}, f<f'\\
    & T^{(f,f')} \geq  -Y^{(f,f')}\times U, \forall f,f'\in \mathcal{F}, f<f'\\
    & T^{(f,f')} \geq 1- Y^{(f,f')}\times (N+1)\times U, \forall f,f'\in \mathcal{F}, f<f'
\end{align}
$Y$ is another intermediate variable used along with $T$ to ensure there is no repeated path. $Y$ is a binary variable and it does not carry a meaningful value. This variable ensures the value assigned to comparison of $f$ and $f'$ stay in the acceptable area.

\subsection{Heuristic Algorithm for Path $\&$ Flow Selection}\label{subsec:heuristic_alg}
The execution time of the mathematical formulation discussed in the previous section for large-scale networks is too high (in the range of several hours or days with respect to the size of the network). Therefore, a heuristic algorithm is proposed in Alg.~\ref{alg:HILP} to solve the corresponding optimization problem in a rational time. The heuristic algorithm proposed for Path $\And$ Flow Selection is called \textit{PFS}. The algorithm is designed in a recursive and greedy manner with a sufficiently low computational complexity.

\begin{algorithm}[!htbp]
	\caption{Pseudo-Code of \textit{path $\&$ flow selection} (PFS)}
	\label{alg:HILP}
	\small
	\allowdisplaybreaks
	\begin{algorithmic}[1]
    	\INPUT{$<$\textit{Topo}, \textit{AL}, \textit{TL}$>$}
    	\LineComment{Topo: Network topology}
    	\LineComment{AL: Allowed lengths for monitoring flows}
    	\Statex{\texttt{TL: Targeted links (links to monitor)}}
    	\OUTPUT{$<$\textit{ULSF},\:\textit{SRSF}$>$}
    	\State{\textit{ULSF}=$\emptyset$}
    	\LineComment{~~ULSF: Used Links So Far}
    	\State{\textit{SRSF}=$\emptyset$}
    	\LineComment{~~SRSF: Selected Routes So Far}
    	\State{\textit{PS} = Extract\_List\_of\_Possible\_Sources(Topo)}
    	\LineComment{~~PS: Possible sources}
    	\State{\textit{A}, \textit{B} = Convert\_Topo\_to\_Graph(\textit{Topo})}
    	\LineComment{~~Mapping IP/MAC to numbers and demonstr-}
    	\LineComment{~~ating as matrix}
    	\LineComment{~~A: Adjacency matrix, B: Capacity matrix}
    	\For{\textit{length} $\in$ \textit{AL}}
        	\For{\textit{src} $\in$ \textit{PS}}
        	    \State{Find\_Routes(\textit{src, length, A, B, *ULSF, *SRSF, TL})}
        	    \LineComment{~~A recursive function calculating }
        	    \LineComment{~~the routes and deciding on the }
        	    \LineComment{~~number of required monitoring flows}
        	\EndFor
    	\EndFor\\
	    \Return{$<$\textit{ULSF,\:SRSF}$>$}
	\end{algorithmic}
\end{algorithm}

\textit{PFS} takes the network topology, targeted links (links to be monitored), and allowed lengths for monitoring flows and makes a decision on the number of required flows. Thereafter, the algorithm computes the routes for every monitoring flow in a way that all targeted links are traversed by at least $x$ flows. In the first 4 lines of Alg.~\ref{alg:HILP}, the preliminaries are done which includes finding the possible sources for the monitoring flows as well as converting the network topology into adjacency and bandwidth matrices. After that, for each possible pair of length of flow and source node, \textit{PFS} generate a couple of flows. This will continue until all possible flows are added, however, one may decide to continue the process until all required links are included in the routes. Having all possible flows, increases both the solution accuracy and the execution time of \textit{DVC}. On the other hand, continuing the process until all required links are included reduces the execution time of \textit{DVC} in exchange for a tiny reduction in the solution accuracy.

\begin{algorithm}[!htbp]
	\caption{Pseudo-Code of Find\_Routes Function}
	\label{alg:FindRoutes}
	\small
	\allowdisplaybreaks
	\begin{algorithmic}[1]
    	\INPUT{$<$\textit{CN, RH, A, B, ULSF, SRSF, TL, DN=$\emptyset$, SR=$\emptyset$, SN=$\emptyset>$}}
    	\LineComment{CN: Current node, RH: Remaining hops}
    	\LineComment{A: Adjacency matrix, B: Capacity matrix}
    	\LineComment{ULSF: Used Links So Far}
    	\LineComment{SRSF: Selected Routes So Far}
    	\LineComment{DN: Destination node, SR: Selected route}
    	\LineComment{SN: Source node, TL: Targeted links}
    	\OUTPUT{$<>$}
    	\If{\textit{RH}=0}
    	    \If{\textit{CN = DN} $\And$ \textit{SR} $\neq\emptyset$}
    	       \State{Update \textit{ULSF} $\And$ \textit{SRSF} based on \textit{SR} and \textit{TL}}
    	    \EndIf
    	\Else
    	    \If{\textit{DN}=$\emptyset$} \textit{DN=CN} \EndIf
    	    \If{\textit{SN}=$\emptyset$} \textit{SN=CN} \EndIf
    	    \For{all nodes \textit{n} connected to \textit{CN}}
    	        \If{\textit{n} $\notin$ \textit{SR} $\And$ no loop in (\textit{SR+n})}
    	            \State{Find\_Routes(\textit{n, RH-1, A, B, ULSF, SRSF, }}
    	            \Statex{~~~~~~~~~~~~~~~~~\textit{DN, SR+n, SN})}
    	        \EndIf
    	    \EndFor
    	\EndIf
	\end{algorithmic}
\end{algorithm}

The pseudo code explained in Alg.~\ref{alg:FindRoutes}, is a recursive approach to find possible monitoring flows which traverse some of the targeted links. The algorithm takes the source, destination, current node, set of targeted links, allowed remained hops (allowed length minus taken hops) as input and returns the set of possible monitoring flows starting from source and ending to destination with specified number of hops.

\subsection{Injection of Flows}\label{subsec:injection_of_flows}
In order to inject the monitoring flows into the network, a component called \textit{Traffic Generator} is defined. This component consists of two main parts: Manager and Agent. The task of the \textit{Traffic Generator Manager (TG Manager)} is to communicate directly with \textit{Path $\&$ Flow Selector} to receive the list of required monitoring flows. Thereafter, \textit{TG Manager} communicates with \textit{Traffic Generator Agents (TG Agents)} and configures them to inject the monitoring flows into the network. Every \textit{TG Agent} instance should inject the requested flows and then returns the end-to-end delay assigned to those flows to \textit{TG Manager}. Thereafter, these values are consolidated and reported to \textit{DVC} module (Fig.~\ref{fig:Architecture}). 

On the other hand, the routing devices (which are SDN switches here) should be configured properly to steer the monitoring flows through the planned routes. To this end, \textit{Forwarding Information Manager (FIM)} module is considered. This module can either be implemented as a part of the SDN controller or as an application out of the SDN controller. Flip-side, the network topology information can be found using \textit{Topology Discovery} module. Similar to \textit{FIM} this module can either be implemented as a part of the SDN controller or as an application out of the SDN controller.

\subsection{Meta-Heuristic Algorithm for Inferring Delay Vector}\label{subsec:pso}
In this subsection, the problem is to take a list of end-to-end delays and routing paths (traversed by each monitoring flow) as input and infer the link delay on targeted links. In fact, the problem is a linear system of equations where the delays of the links are unknowns and the end-to-end delays are fixed values. To solve the aforementioned problem a population swarm optimization (PSO) algorithm is exploited. This meta-heuristic algorithm is called \textit{Delay Vector Calculator (DVC)} algorithm. 
\begin{figure}
	\begin{center}
		\begin{tabular}{|c|c|c|c|c|c|c|c|c|c|c|}\hline
			1.2 & 1.0 & 0.5 & 0.8 & 2 & 1.5 & 0.4 & 0.3 & 0.8 & 1.1\\\hline
		\end{tabular}
		\caption{A particle in the proposed PSO algorithm showing the delay of 10 links. In this example, the measurement unit for all values is considered to be millisecond.}
		\label{fig:particle}
	\end{center}
\end{figure}
In \textit{DVC}, each particle is a list of delays for targeted links. Let the number of targeted links be 10 (considering each direction as one targeted link), Fig.~\ref{fig:particle} shows a sample particle of the proposed algorithm. The first value in the representation of the particle is 1.2 meaning that the link delay for the first targeted link is 1.2ms (considering milliseconds as the measurement unit).

\begin{algorithm}
	\caption{Pseudo-Code of \textit{DVC} Function}
	\label{alg:DVC}
	\small
	\allowdisplaybreaks
	\begin{algorithmic}[1]
    	\INPUT{$<$\textit{A, SR, EED, TL}$>$}
    	\LineComment{A: Adjacency matrix, SR: Selected route}
    	\LineComment{EED: End-to-End delays, TL: Targeted links}
    	\OUTPUT{$<D>$}
    	\State{\textit{population} = Initial\_Population(\textit{TL})}
    	\LineComment{Generate \textit{x-1} particles. For each particle, generate y random value assigning them as a delay value to every link.}
    	\State{\textit{population} += LSS(\textit{EED, SR})}
    	\LineComment{Add least-squares solution of the linear equations as a particle to the population.}
    	\For{t=1: maximum\_generations}
    	    \For{\textbf{each} particle \textit{p} in \textit{population}}
    	        \State{\textit{fp} = Fitness\_Function(\textit{p})}
    	        \LineComment{Use \textit{p} to solve the linear system} 
    	        \LineComment{and consider the fitness as -error.}
    	        \If{\textit{fp} is better than \textit{fpBest}}
    	            \State{\textit{fbest\_p = fp}}
    	            \State{\textit{best\_p = p}}
    	        \EndIf
    	    \EndFor
    	    \For{\textbf{each} particle \textit{p} in \textit{population}}
    	   		\LineComment{Update velocity}
    	   		\For{$i = 1: y$}
    	   		    \State{Generate two random values $r_1$ and $r_2$}
    	   		    \State{$cognitive=c_1\times r_1\times(best\_p[i]-p[i])$}
    	   		    \LineComment{~$c_1$: Cognitive constant (attraction} 
    	   		    \LineComment{~to best particle vs selfishness)}
    	   		    \State{$social=c_2\times r_2\times(best\_p[i]-p[i])$}
    	   		    \LineComment{~$c_2$: Social constant (attraction to}
    	   		    \LineComment{~social behaviour)}
    	   		    \State{$velocity_{p[i]}=w\times velocity_{p[i]}+cognitive+social$}
    	   		    \LineComment{~$w$: Constant inertia weight (how} 
    	   		    \LineComment{~much to weigh the previous}
    	   		    \LineComment{~velocity)}
    	   		\LineComment{Update position}
    	   		\State{$position_{p[i]}=position_{p[i]}+velocity_{p[i]}$}
    	   		\State{Correct $position_{p[i]}$ by keeping it in boundaries}
    	   		\EndFor
    	    \EndFor
    	    \If{fbest\_p is zero}\State{\Return{best\_p}}\EndIf
    	    \If{best\_p has no improvement during \textit{m} generations}
    	        \State{\textit{population} = Mutation(\textit{population})}
    	        \LineComment{~Randomly change some of particles} 
    	        \LineComment{~velocities $\&$ positions}
    	    \EndIf    
    	\EndFor
	\end{algorithmic}
\end{algorithm}

Alg.~\ref{alg:DVC} shows the pseudo code of \textit{DVC}. In this algorithm, a random initial population is generated in line 1 and thereafter, the population is enhanced by adding an elite particle. Let \textit{x} be number of particles and \textit{y} the number of targeted links (unknowns). In order to generate the initial population, $x-1$ particles should be generated. Each particle is a list of \textit{y} values showing the delay of one of the targeted links. Initial\_Population function generates $(x-1)\times y$ random values as the first generation of outcomes. LSS function in line 2 of the algorithm, returns the least-squares solution to a linear matrix equation. LSS solves the equation $ax = b$ by computing a vector $x$ that minimizes the squared Euclidean 2-norm $\| b - a x \|^2_2$. The equation may be under-, well-, or over-determined (i.e., the number of linearly independent rows of $a$ can be less than, equal to, or greater than its number of linearly independent columns)\cite{lstsq}. Besides, due to measurement error is linear system of equations may not have a exact solution, therefore, a PSO algorithm is exploited to find the best possible solution.

Let maximum\_generations be the maximum number of iteration the particles move they positions. Line 3 makes a loop over actions applied to the population to move into a new generation. Thereafter, for each particle in the population, the fitness is calculated. The error of solving the linear systems using the positions reported in a particle is considered as the negative of fitness for that particle. During the calculation of fitness, the best particle is specified. At this point, the particles start to move around considering several factors: their current position, the position of the best particle, and their velocity. Every particle makes a trade-off between going toward the way it is going and approaching the position of the best particle (lines 12-19). If the best particle does not show any improvement in the fitness over \textit{m} generations, this means that either the population is trapped in a local optimum or the measurement error is too high making the linear systems unsolvable. Therefore, to make sure the population is not trapped in a local optimum a mutation algorithm is considered to free some particles from local optimum. Finally, in order to reduce the execution time when the algorithm has found the desired solution, lines 21-23 are developed which stops the algorithm and returns back the best solution found so far.

\section{Implementation}\label{sec:implementation}
The proposed architecture consists of six modules: \textit{Topology Discovery (TD)}, \textit{Forwarding Information Manager (FIM)}, \textit{Path $\&$ Flow Selector (PFS)}, \textit{Delay Vector Calculator (DVC)}, \textit{TG Manager (TGM)}, and \textit{TG Agents (TGA)}. The modules \textit{PFS} and \textit{DVC} are implemented in python and the algorithms are precisely described in sections~\ref{subsec:pso} and~\ref{subsec:heuristic_alg}.
To have a better understanding of the implementation aspects, a brief overview of inputs/outputs of different steps of the proposed architecture is provided below (more details in section~\ref{subsec:Architecture}):
\begin{enumerate}
   \item \textit{PFS} defines monitoring flows and specifies their paths, i.e., input: network topology, output: set of routes, invokes: \textit{TG Manager} and \textit{FIM};
   \item \textit{FIM} enforces flow entries to network switches, i.e., input: set of routes, no output, invokes: nothing;
   \item \textit{TG Manager} finds the E2E delays of all flows in a distributed manner (using \textit{TG Agent}), i.e., input: flows identifiers (e.g., source, destination, and port), output: delay of monitoring flows, invokes: \textit{DVC};
   \item \textit{TG Agent} generates monitoring flows and injects them to the network, i.e., input: flows identifiers (e.g, source and destinations), output: delay of set of (source, destination), invokes: nothing;
   \item \textit{DVC} input: E2E delays, output: Delay Vector, invokes: nothing. 
 \end{enumerate}
In this section the focus is on \textit{Topology Detector} module, \textit{FIM} Module, and \textit{TG} Module. 

\subsection{Topology Detector}\label{subsec:impl_td}
Topology information can be obtained through SDN controller using an API. This information is actually about switches Data-Path ID's (DPID's), hosts MAC, IP addresses, and how these elements are connected together using links. To support cross-platform compatibility, the \textit{Topology Detector} is decided to be implemented out of controller in application layer by utilizing REST API (Fig.~\ref{fig:Architecture}).

\subsection{\textit{FIM} module and TG module}\label{subsec:impl_fim}
According to \ref{subsec:pso}, \textit{DVC} module requires end-to-end delay of selected paths as input. In order to calculate end-to-end delays, we inject monitoring flows to the network and by calculating round trip time of each monitoring flow, we can determine all chosen paths end-to-end delays.
For this purpose, two important tasks are: 1) setting up flow entries and 2) generating monitoring flows. \textit{FIM} module is responsible to get list of paths from \textit{PFS} module and enforce the flow entries; and \textit{TG} module gets list of paths from \textit{PFS} module and then generates monitoring flows to calculate end-to-end delays; 

\subsubsection{How to handle flows with similar source and destination but different paths?}\label{subsec:impl_sc_1}
\extended{
As it is described in Section~\ref{subsec:Architecture}, we use end-to-end delays of selected paths to infer link delay vector. To this end, each \textit{TG Agent} sends ICMP echo request packets to the destinations. Consequently, the destination will answer to this packet using ICMP echo reply. We refer to these ICMP packets as a \textit{monitoring flow}.
\begin{figure}
    \centering
    \includegraphics[width=0.8\columnwidth]{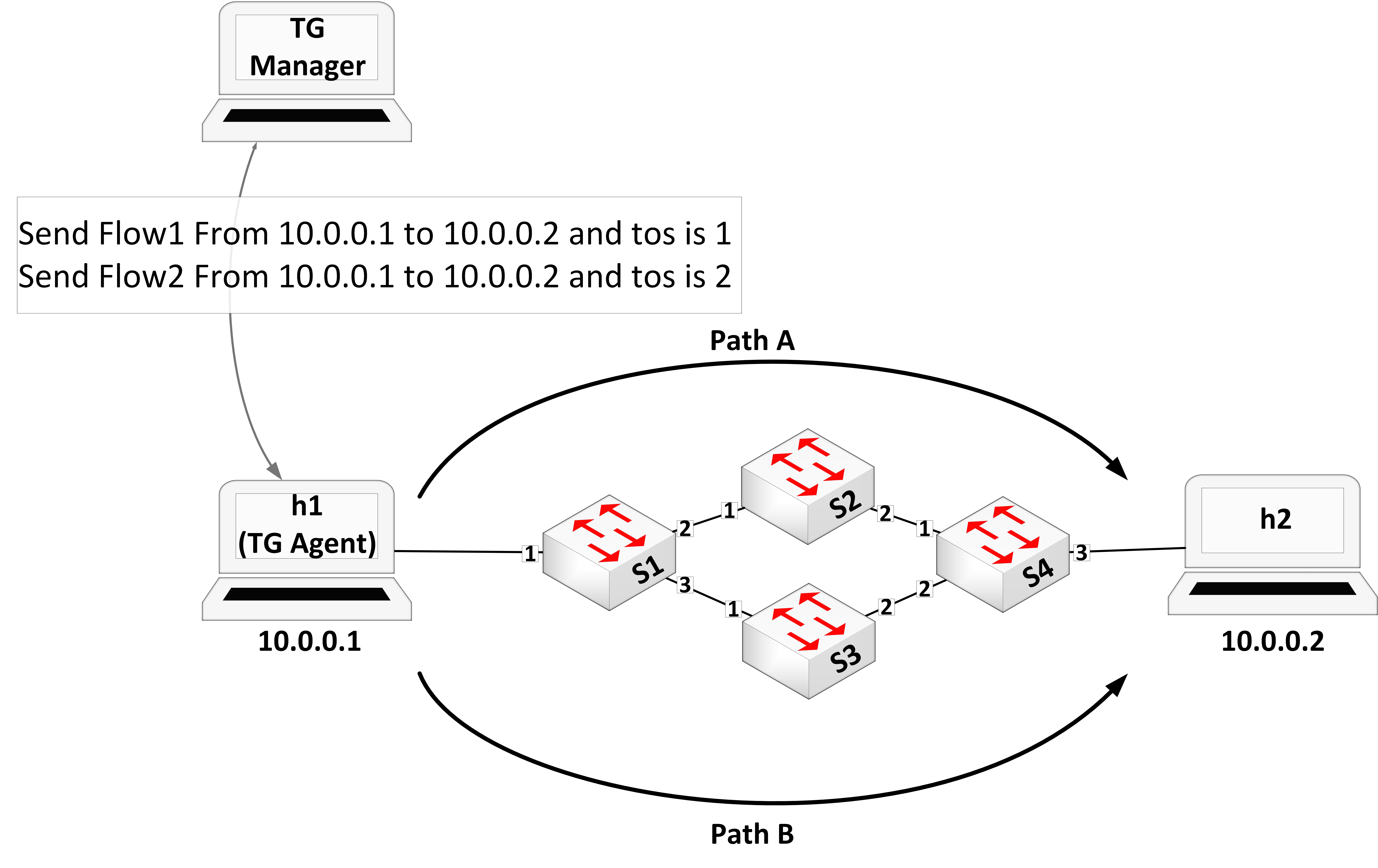}
    \caption{Two flows with similar source and destination, however, traversing via two different paths.}
    \label{fig:Injection of flows 1}
\end{figure}
Fig.~\ref{fig:Injection of flows 1} shows a scenario in which there are two different paths between h1 and h2, the \textit{TG Agent} hosts. These paths are outputs of \textit{PFS} module.
Consider the paths as the following:
\begin{itemize}
    \item Path A: $h1\rightarrow s1\rightarrow s2\rightarrow s4\rightarrow h2$
    \item Path B: $h1\rightarrow s1\rightarrow s3\rightarrow s4\rightarrow h2$
\end{itemize}
Both paths has same pair of source and destination ($h1\rightarrow h2$). 
To make a difference between these two monitoring flows, a tagging mechanism is exploited. 
}
\shortver{We exploit a tagging mechanism to make difference between flows with similar source and destination but different paths (for more explanation refer to \cite{extendedversionofOptimalEstima2020})}
\extended{\textbf{Monitoring Flows Tagging Mechanism:}}
Consider, there are two monitoring flows, with similar source and destination ($h1\rightarrow h2$), traversing through different paths. Since source and destination IPs are similar in both flows, we need to add a unique tag to make a difference between these two flows.
We use the Type of Service (ToS) field in IP header to tag these flows. 
Since ip\_tos is an 8-bit field and each monitoring flow is recognized by ip\_src, ip\_dst, and ip\_tos; Therefore, practically there are 256 monitoring flows available per each source and destination TG Agent pair. 
It is worth mentioning that we use ICMP echo request and ICMP echo reply to estimate round-trip time. Here, round-trip time is the time that takes a monitoring flow to reach the destination and get back to its source (end-to-end delay).

Assigning Monitoring Flows to Paths: 
A \textit{TG Agent} (h1 in this scenario) sends an ICMP echo request packet with ip\_tos=1 (monitoring flow 1) and another one with ip\_tos=2 (monitoring flow 2), respectively. Since reply packets have ip\_tos field similar to request packets~\cite{rfc1349}, we can recognize that each packet is originally related to which flow. This helps switches to forward the replies according to their ip\_tos. Therefore, monitoring flows with same source and destination pair, could be recognized.
Flow entries of Tab.~\ref{tab:flow_tables_example_1} are installed on switches by \textit{FIM} module using Rest API to control the paths that are traversed by each monitoring flow. Having this table in force, flow 1 will traverse via path A and flow 2 will traverse via path B.


\begin{table}
\caption{Flow entries of switches, scenario 1.}\label{tab:flow_tables_example_1}
\centering
\begin{tabular}{|l|l|l|l||l|}
\hline
switch              & ip\_src  & ip\_dst  & ToS & actions \\ \hline
\multirow{4}{*}{S1} & 10.0.0.1 & 10.0.0.2 & 1   & output=2      \\ \cline{2-5} 
                    & 10.0.0.1 & 10.0.0.2 & 2   & output=3      \\ \cline{2-5} 
                    & 10.0.0.2 & 10.0.0.1 & 1   & output=1      \\ \cline{2-5} 
                    & 10.0.0.2 & 10.0.0.1 & 2   & output=1      \\ \hline
\multirow{2}{*}{S2} & 10.0.0.1 & 10.0.0.2 & 1   & output=2      \\ \cline{2-5} 
                    & 10.0.0.2 & 10.0.0.1 & 1   & output=2      \\ \hline
\multirow{2}{*}{S3} & 10.0.0.1 & 10.0.0.2 & 2   & output=2      \\ \cline{2-5} 
                    & 10.0.0.2 & 10.0.0.1 & 2   & output=1      \\ \hline
\multirow{4}{*}{S4} & 10.0.0.1 & 10.0.0.2 & 1   & output=3      \\ \cline{2-5} 
                    & 10.0.0.1 & 10.0.0.2 & 2   & output=3      \\ \cline{2-5} 
                    & 10.0.0.2 & 10.0.0.1 & 1   & output=1      \\ \cline{2-5} 
                    & 10.0.0.2 & 10.0.0.1 & 2   & output=2      \\ \hline
\end{tabular}
\end{table}

\subsubsection{How to handle loops in the path?}\label{subsec:impl_sc_2}

Based on the nature of PSF algorithm, it is required to have a node as both source and destination of a monitoring flow.
To this end, monitoring flows are sent from one \textit{TG Agent} to itself.

\extended{
\begin{figure}
    \centering
    \includegraphics[width=0.8\columnwidth]{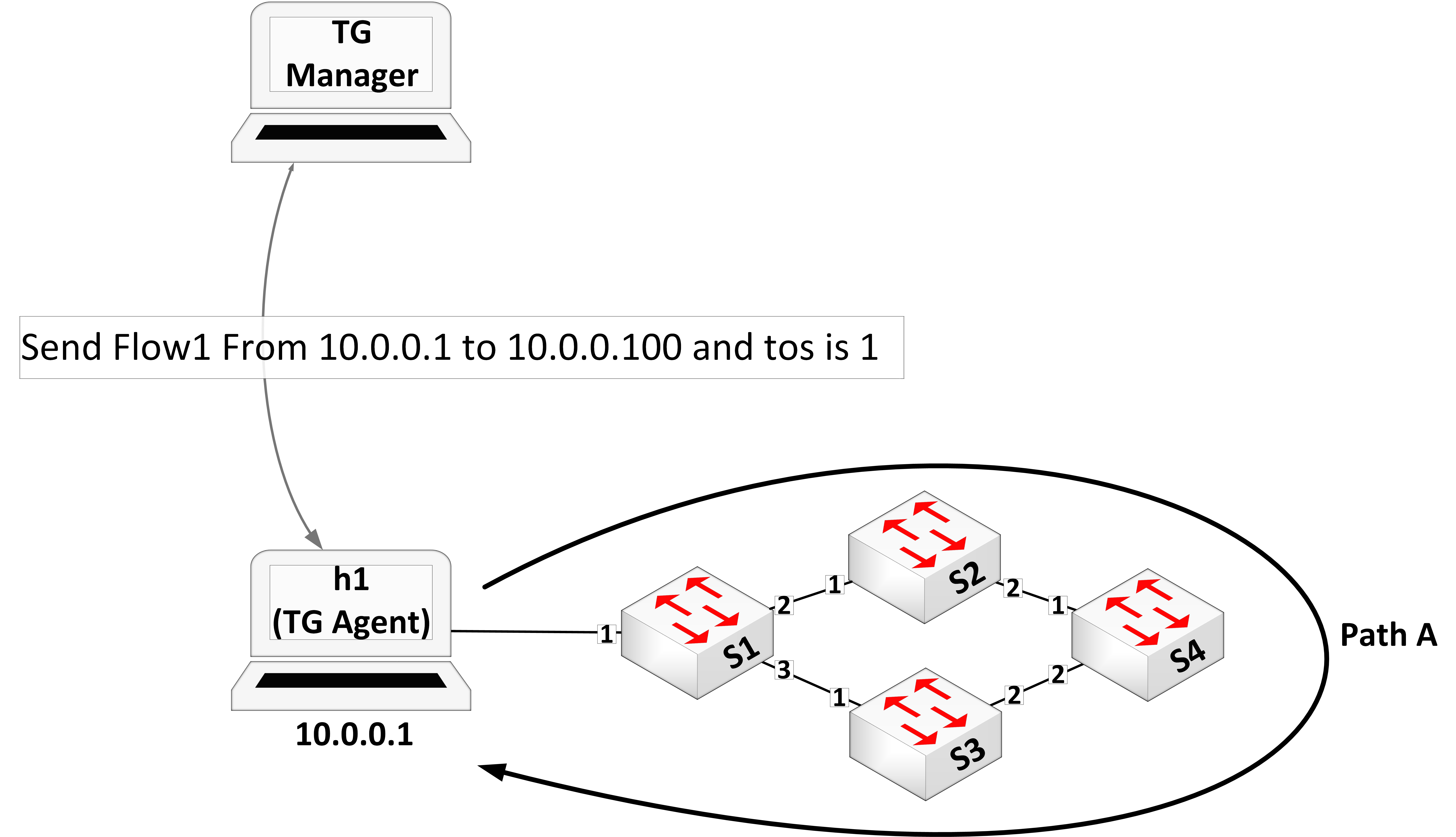}
    \caption{Loop in the path.}
    \label{fig:Injection of flows 2}
\end{figure}
Consider Fig.~\ref{fig:Injection of flows 2}. A monitoring flow must be sent flow h1 to s3 and the reply must get back to the h1. The path will be like the following:}

\extended{
h1$\rightarrow$ s1$\rightarrow$ s2$\rightarrow$ s4$\rightarrow$ s3$\rightarrow$ s1$\rightarrow$ h1}

This situation is more complicated than the former. 
Since the monitoring flow's source and destination are the same, we need to consider a hypothetical host which we consider as this kind of monitoring flows destination.

At h1, a \textit{TG Agent} creates a monitoring flow with the source IP address of h1 and destination IP address of Helper and then send it to the Helper. Like the previous scenario, an ip\_tos is set to tag it (e.g. ip\_tos=1). For instance in s1 switch, we install tab. \ref{tab:emulation_scenario_2_flow_table_1}  flow entry is required.

\begin{table}[ht]
\caption{Scenario 2, sample flow entry with Helper IP's destination that is installed on S1.}
\label{tab:emulation_scenario_2_flow_table_1}
\centering
\begin{tabular}{|c|l|l||l|}
\hline
\multicolumn{3}{|l||}{Match Fields} & \multirow{2}{*}{actions} \\ \cline{1-3} \cline{1-3}
ip\_src  & ip\_dst    & ip\_tos &    \\ \hline
10.0.0.1 & 10.0.0.100 & 1       & output: 2  \\ \hline
\end{tabular}
\end{table}

In other switches, flow entries (\ref{tab:flow_tables_example_2}) are installed in the same way. But the main point is that if we return the same packet produced by h1 to itself, h1 can't accept it as a reply of it's request; Since h1 waits for a packet with the header that is shown in tab. \ref{tab:emulation_scenario_2_packet_header}.
h1 receives a packet that it's source is itself, something like header that is presented in Tab.~\ref{tab:emulation_scenario_2_packet_header2}.
Using a flow entry (see Tab.~\ref{tab:example2_tab1}), we can modify this kind of flows as an acceptable ICMP echo reply by swapping ipv4\_src \& ipv4\_dst, and also changing icmp\_v4\_type from 8 to 0.

Note that in ICMP echo request packets, the icmp\_type field is 8 and in ICMP echo reply packets the icmp\_type field must be 0. 
Therefore, a small modification should happened in the return packet. This change occurs in switch s1 (last switch) using a flow entry which is shown in Tab.~\ref{tab:example2_tab1}. 
Finally, all the flow entries that should be installed on the switches could be seen in Tab.~\ref{tab:flow_tables_example_2}.

\begin{table}[ht]
\caption{Scenario 2, expected ICMP echo reply packet header.}\label{tab:emulation_scenario_2_packet_header}
\centering
\begin{tabular}{|l|l|l|l|}
\hline
ip\_src  & ip\_dst    & ip\_tos & icmp\_type   \\ 
\hline
10.0.0.100 & 10.0.0.1 & 1       & 0  \\
\hline
\end{tabular}
\end{table}

\begin{table}[ht]
\caption{Scenario 2, seen packet header.}\label{tab:emulation_scenario_2_packet_header2}
\centering
\begin{tabular}{|l|l|l|l|}
\hline
ip\_src  & ip\_dst    & ip\_tos & icmp\_type   \\ \hline
10.0.0.1 & 10.0.0.100 & 1       & 8 \\ \hline
\end{tabular}
\end{table}

\begin{table}[ht]
\caption{Scenario 2, switch S1 flow entries for monitoring flows that are traversing back to the same \textit{TG Agent}.}\label{tab:example2_tab1}
\centering
\begin{tabular}{|l|l|l||l|}
\hline
ip\_src  & ip\_dst    & ip\_tos & actions   \\ \hline
10.0.0.1 & 10.0.0.100 & 1       &  
\begin{tabular}[c]{@{}l@{}}output=1,\\ set\_field=ipv4\_src$\rightarrow$10.0.0.100,\\ set\_field=ipv4\_dst$\rightarrow$10.0.0.1,\\ set\_field=icmpv4\_type$\rightarrow$0\end{tabular}  \\ \hline
\end{tabular}
\end{table}

\begin{table}[ht]
\caption{Scenario 2, Flow entries of all switches.}\label{tab:flow_tables_example_2}
\centering
\resizebox{\columnwidth}{!}{%

\begin{tabular}{|l|l|l|l|l||l|}
\hline
switch              & in\_port & ip\_src  & ip\_dst    & ToS & actions                                                                                                                                         \\ \hline
\multirow{2}{*}{S1} & 1        & 10.0.0.1 & 10.0.0.100 & 1   & output=2                                                                                                                                        \\ \cline{2-6} 
                    & 3        & 10.0.0.1 & 10.0.0.100 & 1   & \begin{tabular}[c]{@{}l@{}}output=1,\\ set\_field=ipv4\_src$\rightarrow$10.0.0.100,\\ set\_field=ipv4\_dst$\rightarrow$10.0.0.1,\\ set\_field=icmpv4\_type$\rightarrow$0\end{tabular} \\ \hline
S2                  & 1        & 10.0.0.1 & 10.0.0.100 & 1   & output=2                                                                                                                                        \\ \hline
S3                  & 2        & 10.0.0.1 & 10.0.0.100 & 2   & output=1                                                                                                                                        \\ \hline
S4                  & 1        & 10.0.0.1 & 10.0.0.100 & 1   & output=2                                                                                                                                        \\ \hline
\end{tabular}}
\end{table}




\section{Evaluation}\label{sec:evaluation}
We use Mininet~\cite{kaur2014mininet} as emulation tool, Floodlight as the SDN-controller, and Python as the programming language. The typologies used to evaluate the proposed solution are Abilene and RedIRIS~\cite{topologyzoo}. To run our prototype, we consider there are a few randomly-placed hosts, as monitoring nodes. We also have IPerf generating traffic in the background to make a more realistic scenario.


\subsection{Monitoring Coverage}\label{subsec:eval_coverage}
\begin{figure}[!h]
    \begin{subfigure}{0.49\columnwidth}
       \centering
        \includegraphics[width=\columnwidth]{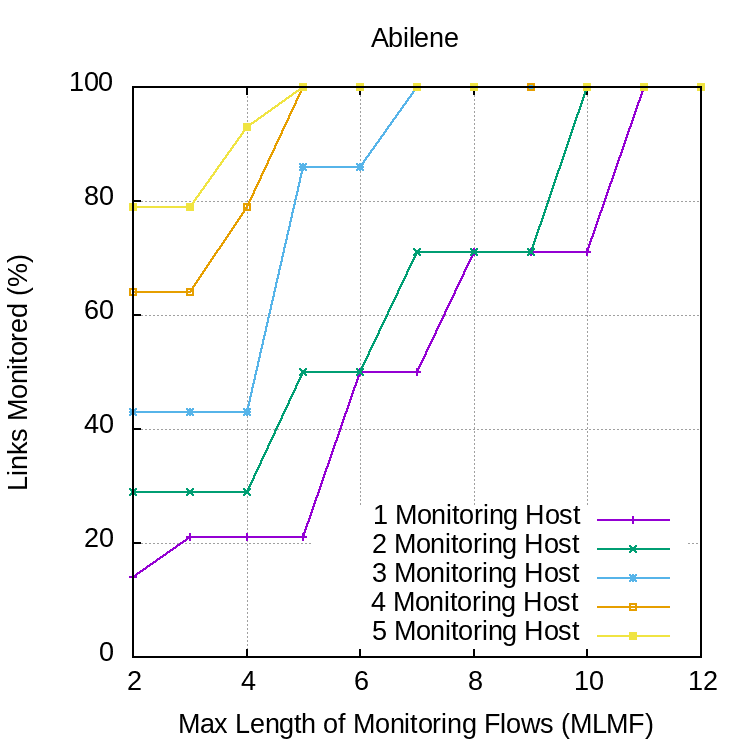}
    \end{subfigure}
    \begin{subfigure}{0.49\columnwidth}
      \centering
      \includegraphics[width=\columnwidth]{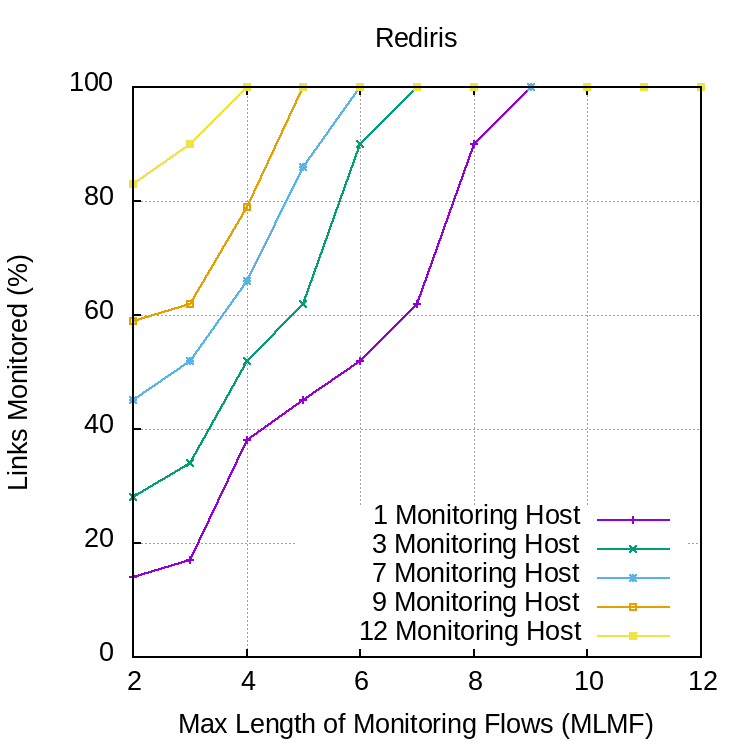}
    \end{subfigure}
    \caption{Monitoring Coverage.}
    \label{fig:Max_Length_of_Routes_Links_Monitored}
\end{figure}

Fig.~\ref{fig:Max_Length_of_Routes_Links_Monitored}, shows the coverage of different scenarios versus the maximum length of monitoring flows (MLMF). Based on the simulation results, there is a direct relationship between MLMF, number of monitoring nodes, and the measurement coverage. Increasing the number of monitoring nodes increases the chance of having full coverage using flows with a lower length. As an example, with five monitoring nodes, all links delays can be measured with flows of length lower than 5 hops (in both topologies); While this value is 12 in Abilene (and 9 in RedIRIS) when there is only one monitoring node.
On the other hand, allowing MLMF to increase, reduces the number of required monitoring nodes to reach full coverage. 

It should be mentioned that having a relatively high MLMF (in this example 11), all links could be monitored with a few number of monitoring nodes (in this example, one monitoring node). However, this is not desirable because increasing MLMF may increase the optimality gap or configuration overhead. On the other hand, increasing the number of monitoring nodes above a number where a relatively low MLMF covers all links, is not necessary. This is because after this point, increasing the number of monitoring flow only increases the complexity of algorithm. 
\begin{figure}[!h]
    \begin{subfigure}{0.49\columnwidth}
       \centering
        \includegraphics[width=\columnwidth]{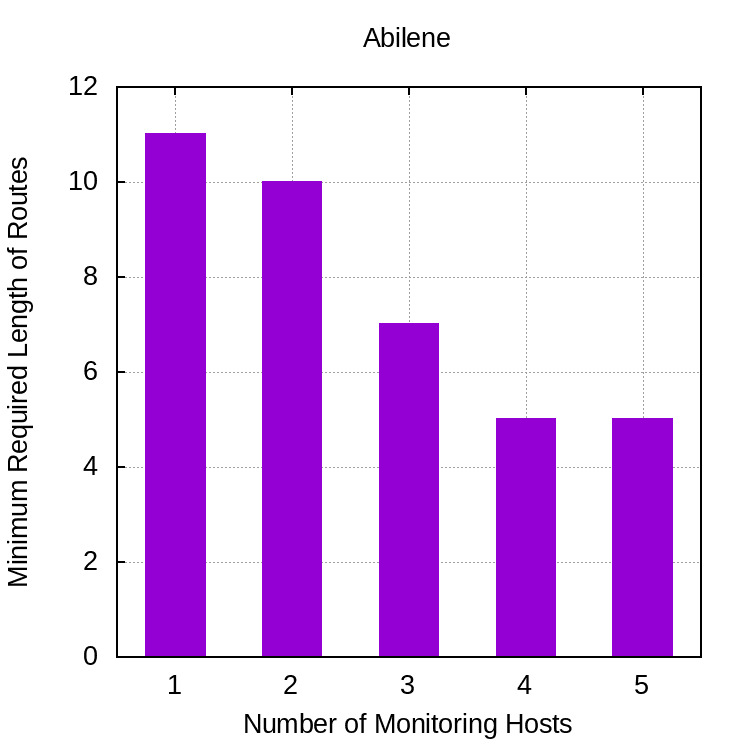}
    \end{subfigure}
    \begin{subfigure}{0.49\columnwidth}
      \centering
      \includegraphics[width=\columnwidth]{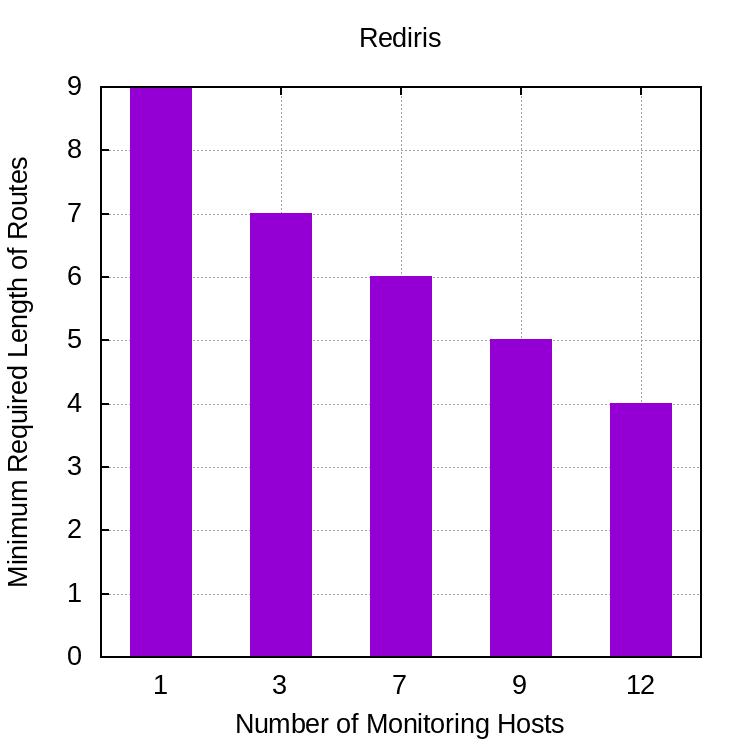}
    \end{subfigure}
    \caption{Minimum MLMF to have full coverage}
    \label{fig:bar_min_MLMF_full_cover}
\end{figure}
Fig.~\ref{fig:bar_min_MLMF_full_cover}, represents the minimum MLMF that can cover all links versus the number of monitoring nodes. The higher the number of monitoring nodes, the lower the minimum of MLMF becomes. This happens because increasing the number of monitoring nodes, reduces the maximum distance of links from monitoring nodes. Consequently, with a lower MLMF, one can monitor the desired links in the network. 

\subsection{Execution Time}\label{subsec:eval_exec_time}
\begin{figure}[!h]
    \begin{subfigure}{0.49\columnwidth}
       \centering
        \includegraphics[width=\columnwidth]{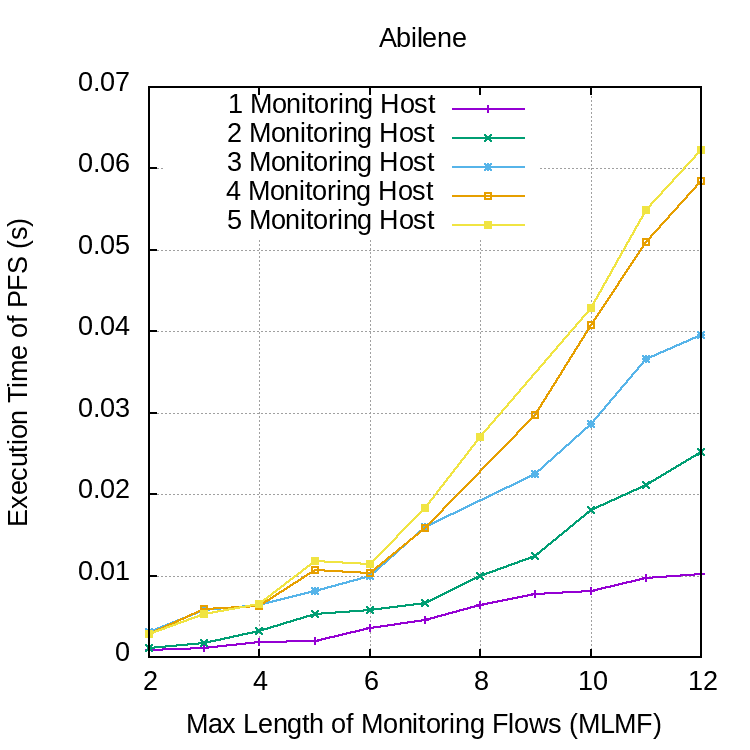}
    \end{subfigure}
    \begin{subfigure}{0.49\columnwidth}
      \centering
      \includegraphics[width=\columnwidth]{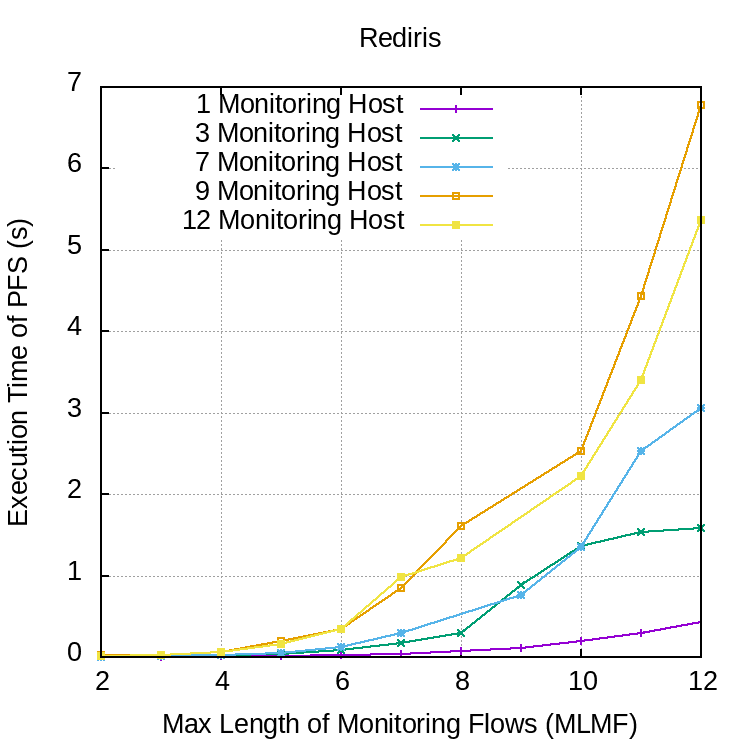}
    \end{subfigure}
    \caption{Execution Time of PFS (a portion of the Offline part)}
    \label{fig:Abilene_time_PFS}
\end{figure}

The execution time of the proposed solution depends on two main components: \textit{PFS} and \textit{DVC}. In the following the execution time of these components is discussed. Fig.~\ref{fig:Abilene_time_PFS}, depicts the execution time of \textit{PFS} versus the maximum allowed length of probe flows. In this plot, five different scenarios of Abilene and RedIRIS topologies (by changing the number of monitoring nodes) are considered. Increasing the possible length of monitoring flows, increases the search space for the algorithm, therefore, increasing the max length of routes increases execution time. On the other hand, increasing the number of monitoring nodes, increases the number of possible combination of monitoring flows. Again, this increases the search area and consequently increases the execution time. It should be considered that decreasing the number of monitoring nodes, may end up to increase the Maximum Length of Monitoring Flows (MLMF). Otherwise, some links will not be covered in the measurement. The execution time of PFS in RedIRIS is significantly higher than Abilene. This happens because the number of nodes in RedIRIS is dramatically higher than Abilene which makes the search space bigger. It should be mentioned that this part is an offline and one-time part. This means that we need to run \textit{PFS} just once and before starting the measurement. Besides, this algorithm is capable of being fully paralleled which makes it significantly faster.

\begin{figure}[!htb]
    \begin{subfigure}{0.49\columnwidth}
       \centering
        \includegraphics[width=\columnwidth]{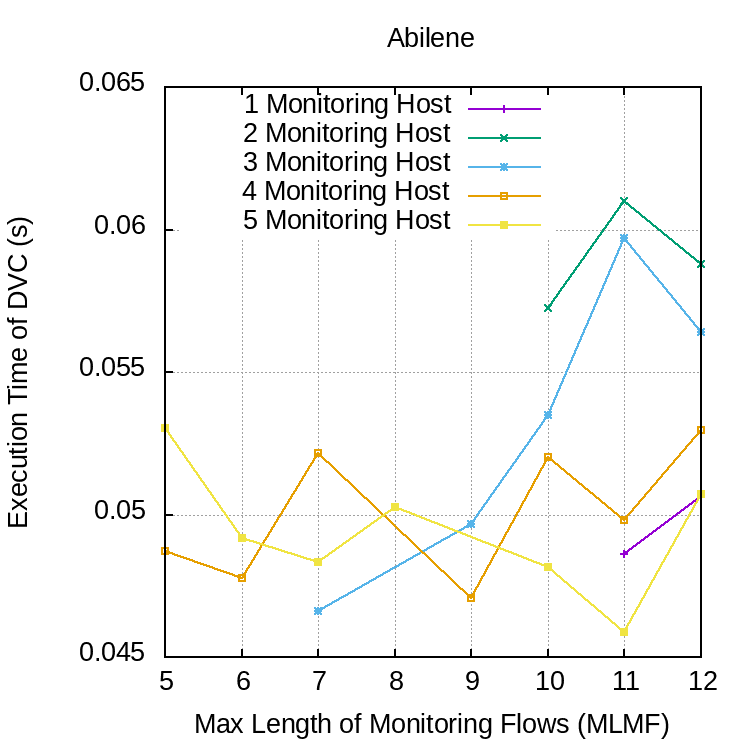}
    \end{subfigure}
    \begin{subfigure}{0.49\columnwidth}
      \centering
      \includegraphics[width=\columnwidth]{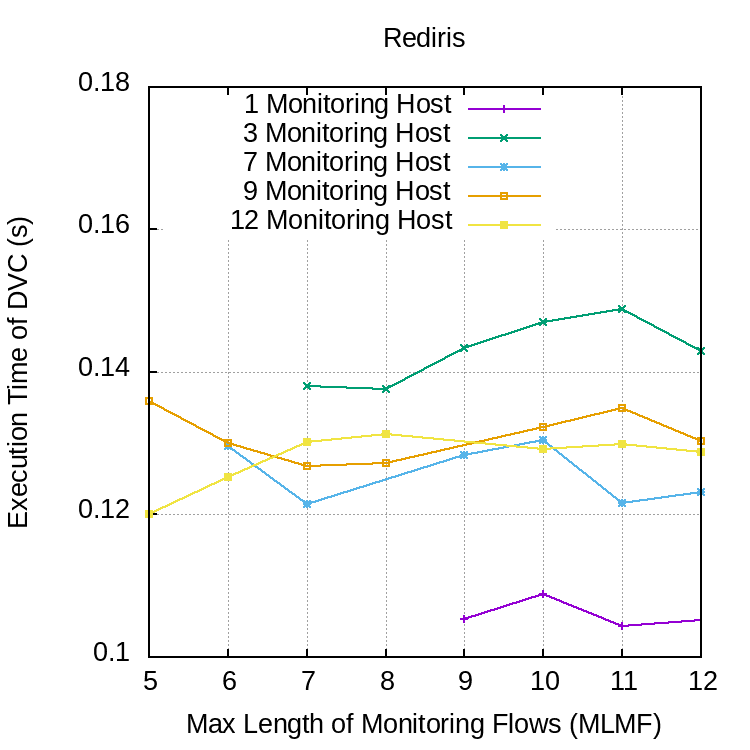}
    \end{subfigure}
    \caption{Execution Time of \textit{DVC} (a portion of the online part)}
    \label{fig:Abilene_time_DVC}
\end{figure}

Fig.~\ref{fig:Abilene_time_DVC} represents the execution time of \textit{DVC} versus the maximum length of monitoring flows (MLMF). Similar to Fig.~\ref{fig:Abilene_time_PFS}, five different scenarios are considered. In this plot, some lines start from the middle of x-axis meaning that the algorithm could not cover all links with an MLMF less than that point. For example, the line showing "2 Monitoring nodes" starts from x=10, this means that at least MLMF should be 10 to cover all links with only 2 monitoring nodes. Since the number of flows is always kept below a threshold, the execution time of DVC is almost the same for different values of MLMF. On the other hand, it is linearly related to the number of investigated links, i.e., the number of routing nodes and the topology.

\begin{figure}[!htb]
    \begin{subfigure}{0.48\columnwidth}
      \centering
      \includegraphics[width=\columnwidth]{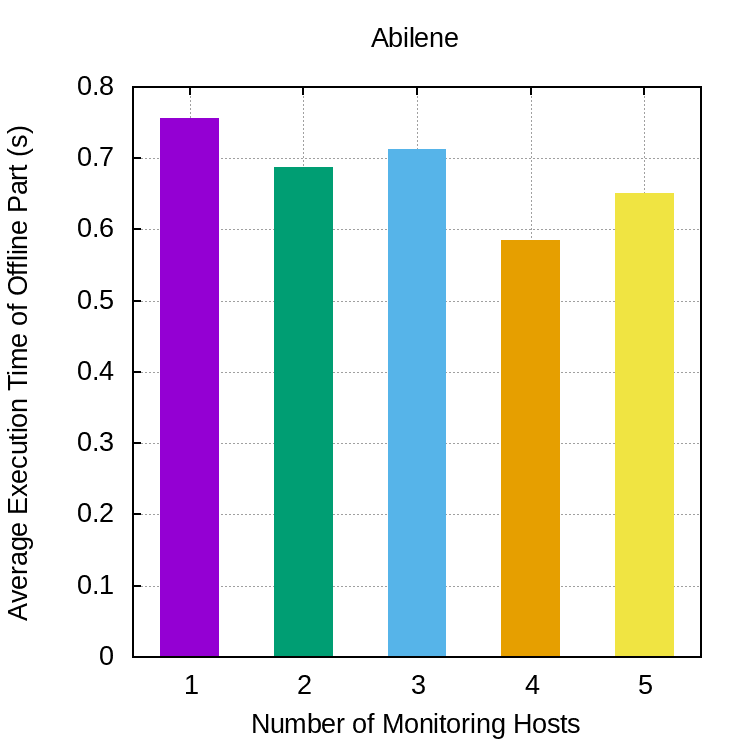}
    \end{subfigure}
    \begin{subfigure}{0.49\columnwidth}
      \centering
      \includegraphics[width=\columnwidth]{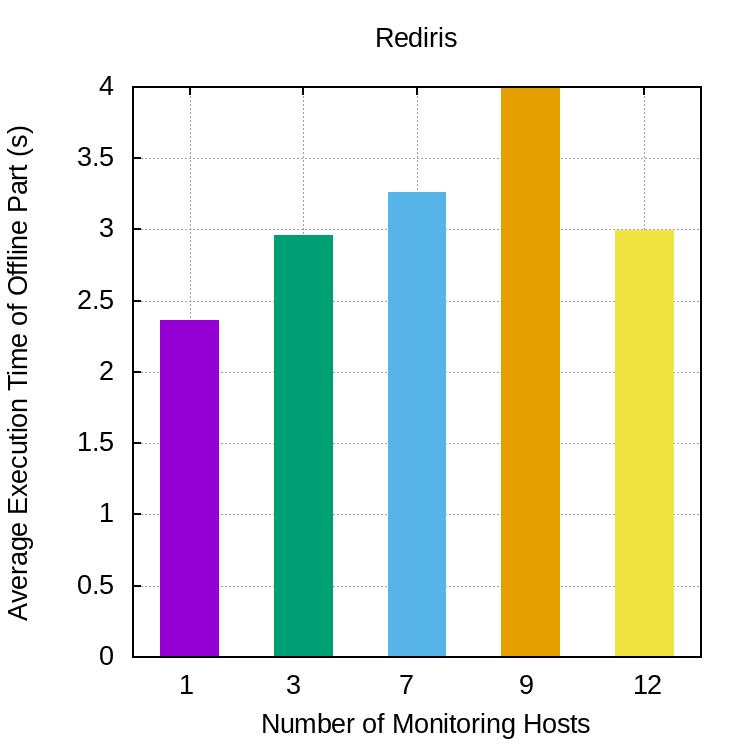}
    \end{subfigure}
    \caption{The total execution time of offline parts.}
    \label{fig:offline_exec_time}
\end{figure}

\begin{figure}[!htb]
    \begin{subfigure}{0.48\columnwidth}
      \centering
      \includegraphics[width=\columnwidth]{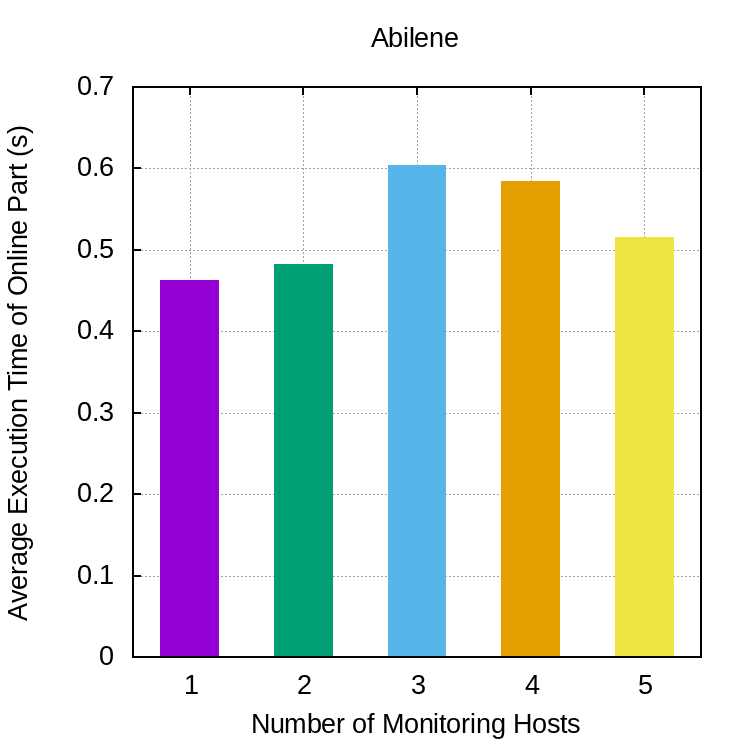}
    \end{subfigure}
    \begin{subfigure}{0.49\columnwidth}
      \centering
      \includegraphics[width=\columnwidth]{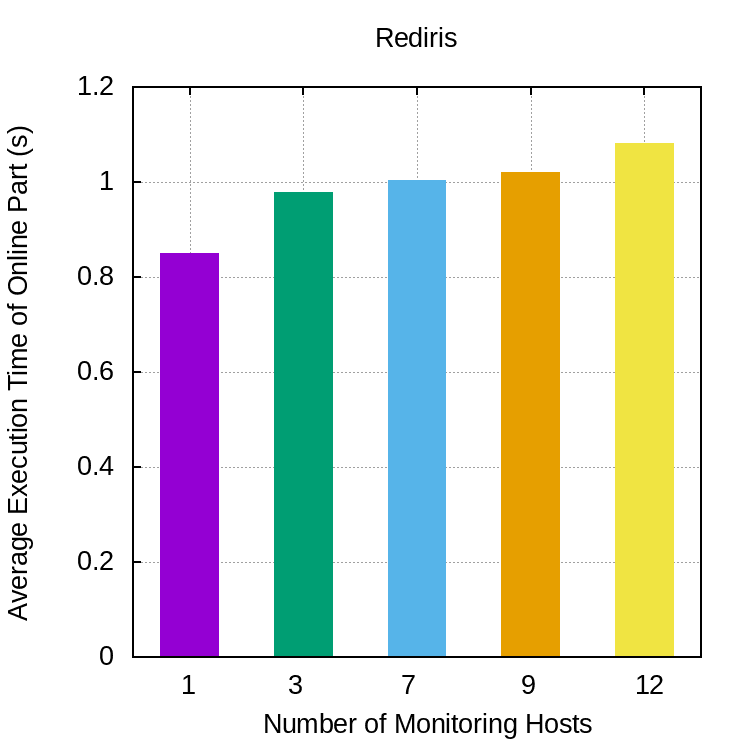}
    \end{subfigure}
    \caption{The total execution time of online parts.}
    \label{fig:online_exec_time}
\end{figure}

To conclude this sub-section, we have plotted the total offline and online execution times for both Abilene and RedIRIS in Figures~\ref{fig:offline_exec_time}~and~\ref{fig:online_exec_time}. The offline part is composed of one-off algorithms that are required to be executed before starting the measurement. \textit{PFS} and rule enforcement in \textit{Traffic Generator} are located in this category. On the other hand, online algorithms should be running continuously to have a real-time LDV measurement. The execution time of the offline part does not impact the measurement time-gap with the real state of links. 
There are two main players in the online part of the proposed solution: \textit{DVC} and \textit{Traffic Generator (TG)}. The majority of time spent in online part is dedicated to \textit{TG}. This happens because our \textit{TG} implementation is sequential and blocking. Therefore, the execution time of \textit{TG} can be reduced significantly by parallelizing probes generation. 

\subsection{Monitoring accuracy and error}\label{subsec:eval_accuracy_and_error}
\begin{figure}
    \begin{subfigure}{0.49\columnwidth}
       \centering
        \includegraphics[width=\columnwidth]{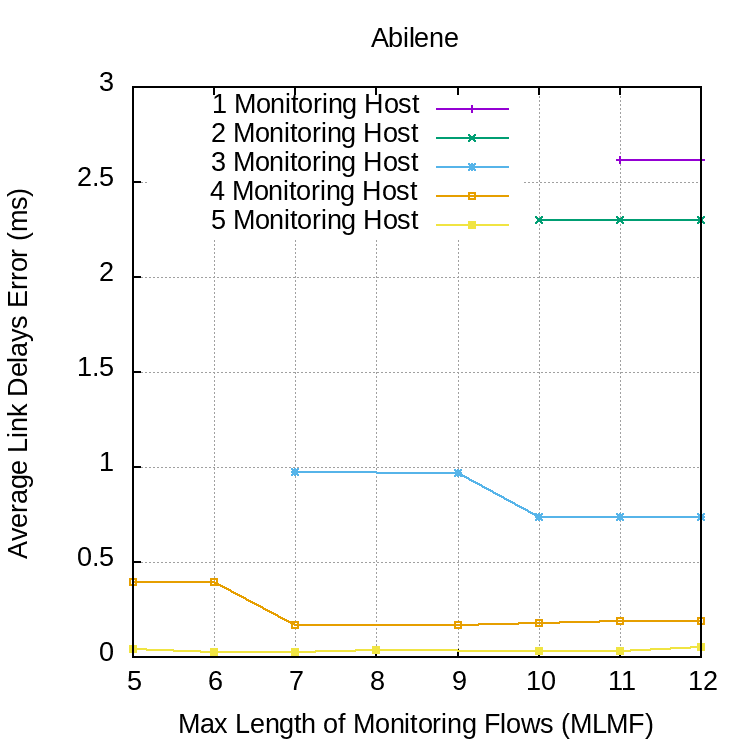}
    \end{subfigure}
    \begin{subfigure}{0.49\columnwidth}
      \centering
      \includegraphics[width=\columnwidth]{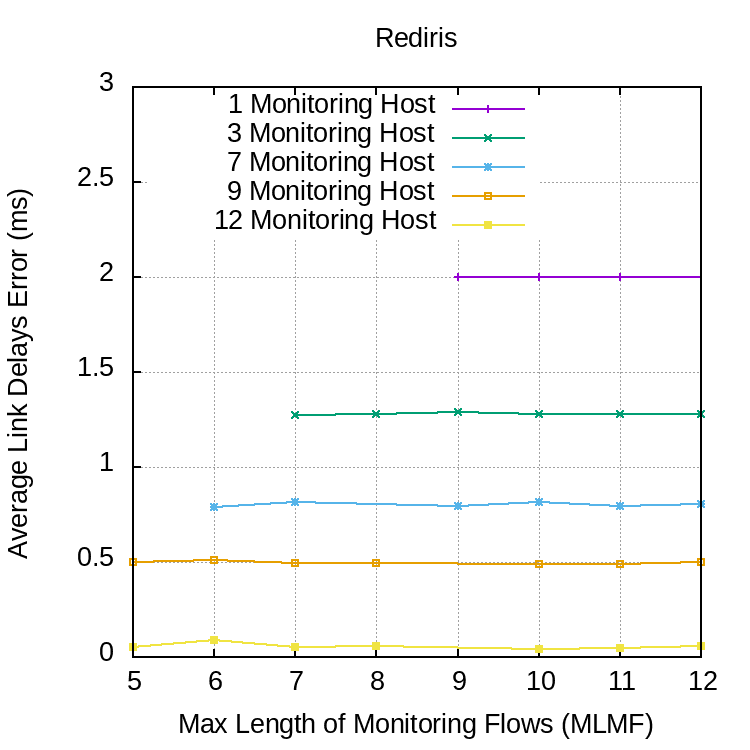}
    \end{subfigure}
    \caption{Average of all link errors.}
    \label{fig:sum_delay_abilene}
\end{figure}

\begin{figure}
\centering
    \begin{subfigure}{0.49\columnwidth}
       \centering
        \includegraphics[width=\columnwidth]{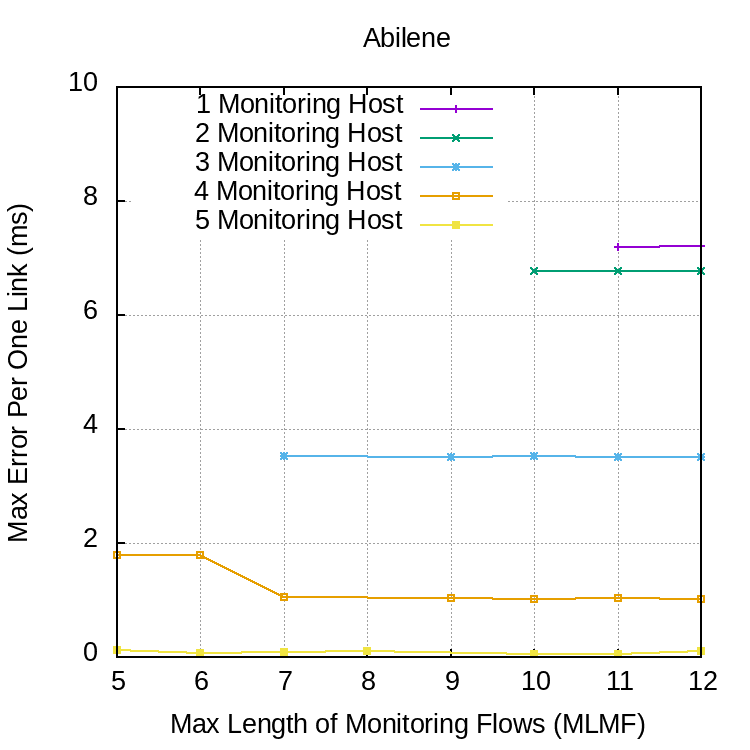}
    \end{subfigure}
    \begin{subfigure}{0.49\columnwidth}
      \centering
      \includegraphics[width=\columnwidth]{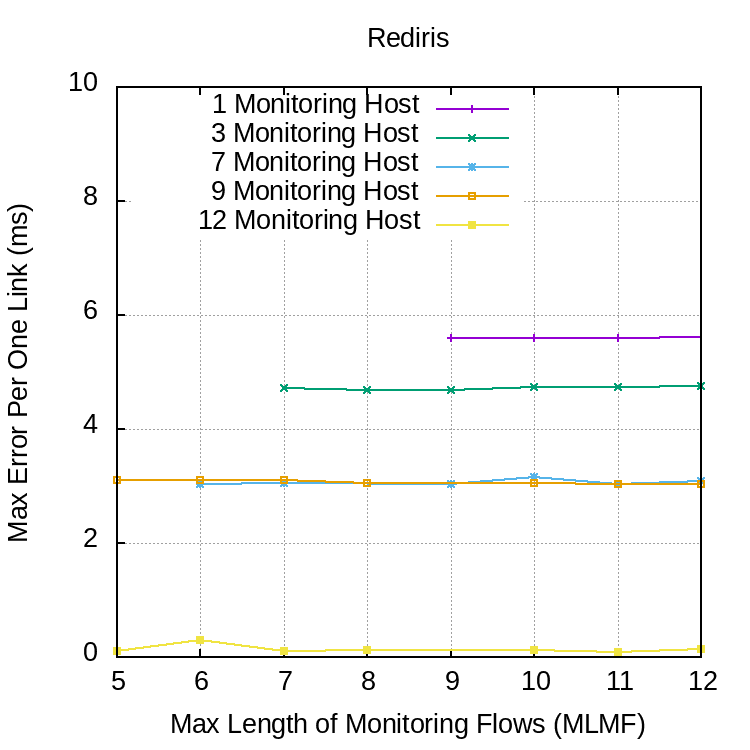}
    \end{subfigure}
    \caption{Max error per link.}
    \label{fig:max_error_abilen}
\end{figure}

\begin{figure}
\centering
    \begin{subfigure}{0.9\columnwidth}
      \centering
      \includegraphics[width=\columnwidth]{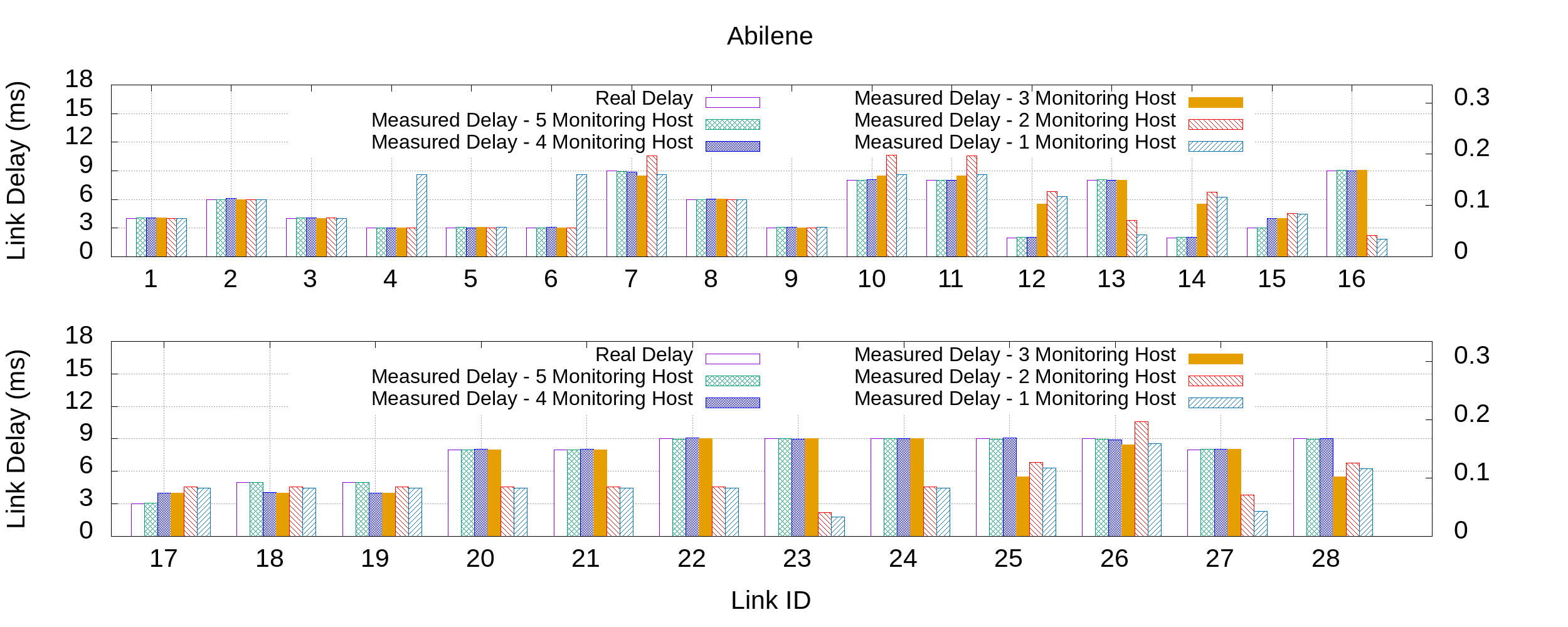}
      \caption{Abilene.}
      \label{fig:real_delay_abile}
    \end{subfigure}
    \begin{subfigure}{0.9\columnwidth}
      \centering
      \includegraphics[width=\columnwidth]{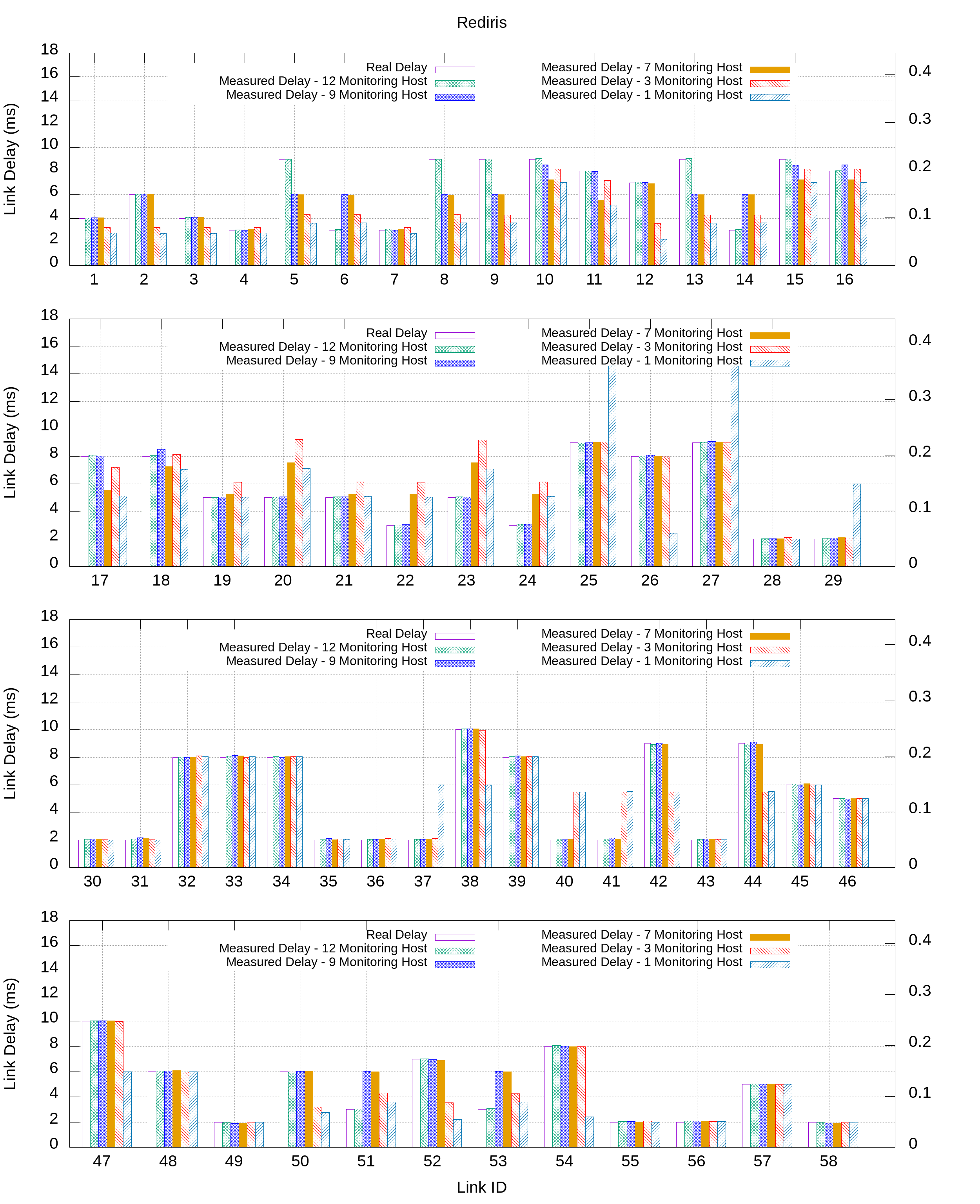}
     \caption{RedIRIS.}
     \label{fig:real_delay_rediris}
    \end{subfigure}
    \caption{Link delay configured in Mininet versus measured link delay.}
    \label{fig:real_delay}
\end{figure}

In this subsection, the monitoring accuracy and the corresponding error is discussed. Fig.~\ref{fig:sum_delay_abilene}, shows the summation of all link delay errors versus MLMF for a different number of monitoring nodes. Some lines do not appear from the beginning of x-axis. This means that before that point, the algorithm could not fully cover all links, i.e., some links are not monitored. This is directly related to the relation of MLMF and number of monitoring nodes. With a low MLMF value, a higher number of monitoring flows should be used to cover all links. On the other hand, increasing the number of monitoring nodes decreases the overall error but in the cost of increasing the monitoring cost (resources).
Similarly, Fig.~\ref{fig:max_error_abilen} presents maximum error among all monitored links versus MLMF considering different number of monitoring nodes. Although increasing MLMF decreases the error, the impact is not significant. To understand this, we should mention that increasing the length of monitoring flows may end up to a higher error rate, however, MLMF is not the length of monitoring flows but the maximum length of monitoring flows. In other words, if MLMF is four then monitoring flows with a length of one, two, three, or four are acceptable. Therefore, increasing MLMF does not increase the error, however, since using a higher length of monitoring flows may end up to a higher measuring error, the impact of increasing MLMF on error rate is not significant.
Fig.~\ref{fig:real_delay}, provides a closer look to per-link delay accuracy with a different number of monitoring flows. Fig.~\ref{fig:real_delay_abile} shows the measured delay using 1-5 monitoring nodes versus the delay configured in Mininet for every link. 
As discussed earlier, the higher the number of monitoring nodes, the better the measurement. It is worth mentioning that although the accuracy of one-monitoring node scenario is higher than the accuracy of two-monitoring nodes scenario in some links, the overall accuracy of two-monitoring nodes is higher. In Fig.~\ref{fig:real_delay_rediris}, measurement accuracy for RedIRIS takes the same pattern as it has for Abilene.

\begin{figure}
    \begin{subfigure}{0.48\columnwidth}
      \centering
      \includegraphics[width=\columnwidth]{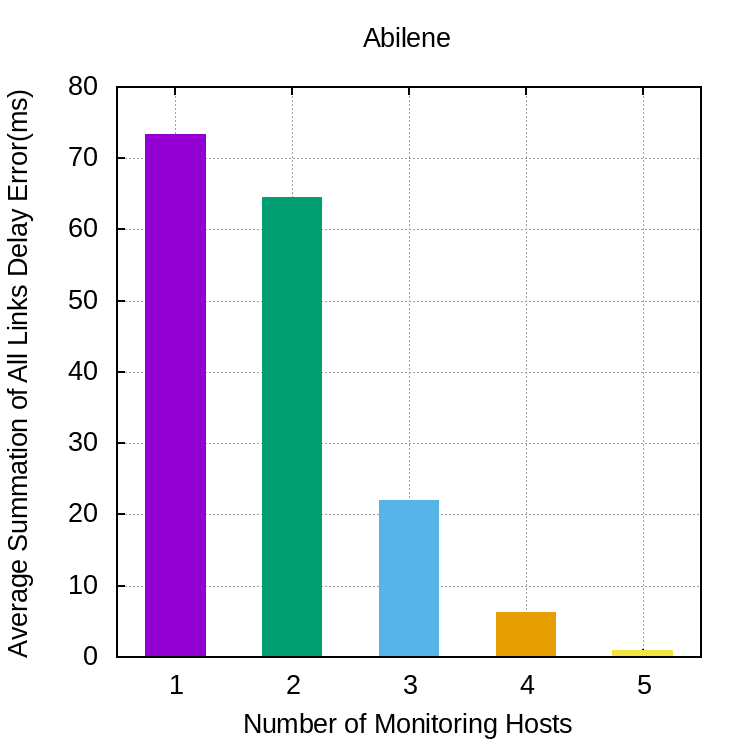}
    \end{subfigure}
    \begin{subfigure}{0.49\columnwidth}
      \centering
      \includegraphics[width=\columnwidth]{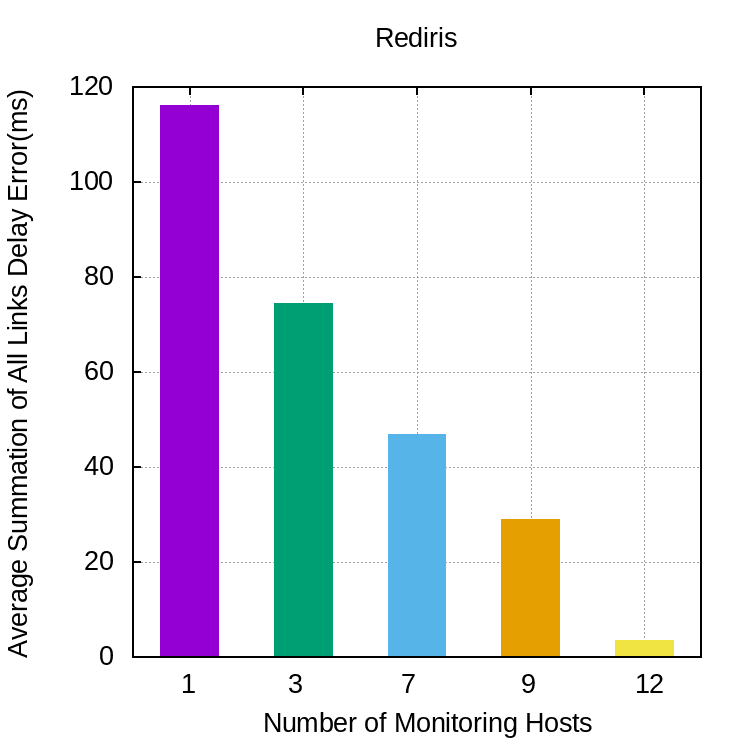}
    \end{subfigure}
    \caption{Impact of the number of monitoring nodes on the link delay error (MLMF=11)}
    \label{fig:eval_Abilene_summation_of_all_links_delay_error_ms_bar_chart}
\end{figure}

Fig.~\ref{fig:eval_Abilene_summation_of_all_links_delay_error_ms_bar_chart} shows the average measurement errors versus the number of monitoring nodes. As the number of monitoring nodes increases the average error is decreases. This happens because the accessibility to different parts of the network increases. In other words, when there is a few number of monitoring nodes, \textit{PFS} cannot define enough monitoring flows. 

\subsection{Networking Overhead}\label{subsec:eval_monitoring_overhead}
\begin{figure}
    \begin{subfigure}{0.48\columnwidth}
      \centering
      \includegraphics[width=\columnwidth]{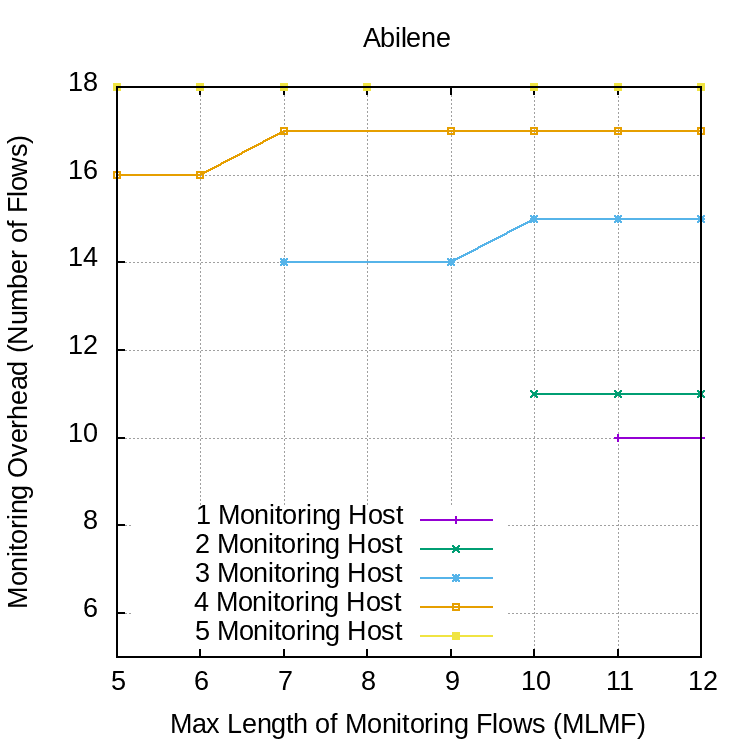}
    \end{subfigure}
    \begin{subfigure}{0.49\columnwidth}
      \centering
      \includegraphics[width=\columnwidth]{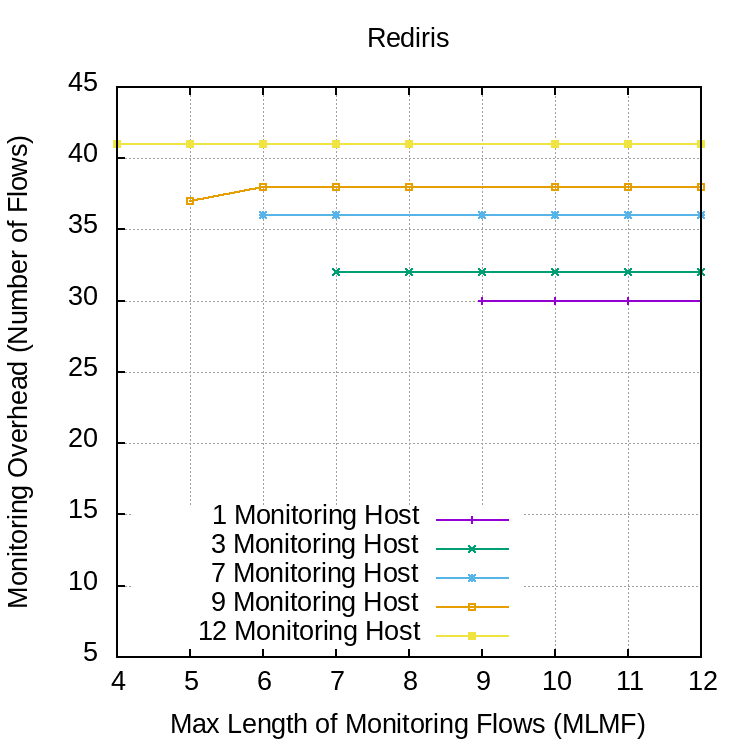}
    \end{subfigure}
    \caption{Monitoring Overhead - Number of monitoring flows}
    \label{fig:required_flows}
\end{figure}

\begin{figure}
    \begin{subfigure}{0.48\columnwidth}
      \centering
      \includegraphics[width=\columnwidth]{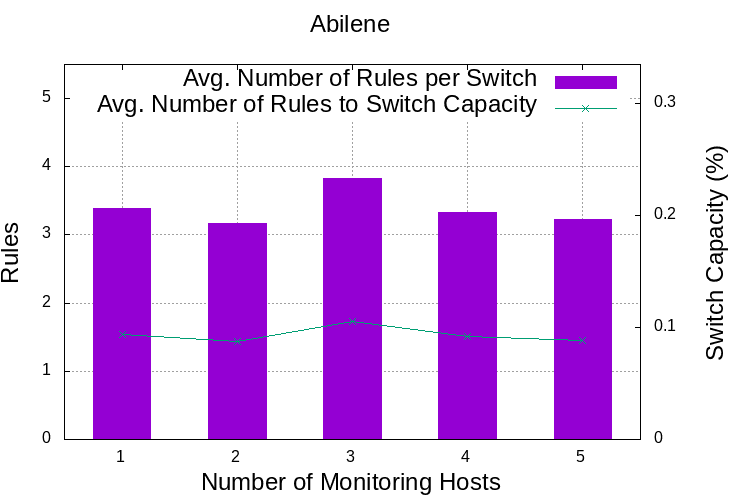}
    \end{subfigure}
    \begin{subfigure}{0.49\columnwidth}
      \centering
      \includegraphics[width=\columnwidth]{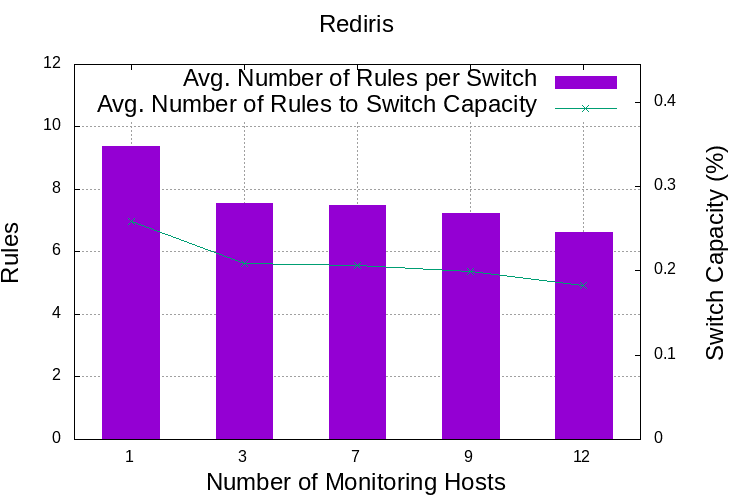}
    \end{subfigure}
    \caption{Monitoring Overhead - Average number of rules per switch, MLMF=11}
    \label{fig:eval_barchart_Average_Percent_of_Rules_to_Switch_Rule_Capacity_Ratio_max_len_of_route_11}
\end{figure}

In order to route monitoring flows, we need to push rules into the switches. Since the length of the switch flow-table is limited, the number of different flows (which impact the number of pushed rules) can be considered as a metric to measure the monitoring overhead. 
Fig.~\ref{fig:required_flows} depicts the number of monitoring flows versus MLMF for different number of monitoring nodes. 
When the number of monitoring nodes increases, the number of monitoring flows increases as well ending up with an increment in the monitoring overhead increases. Similarly, increasing MLMF increases the number of monitoring flows and consequently the monitoring overhead. It is worth mentioning that increasing MLMF beyond a threshold does not impact the overhead. This threshold is the maximum distance between each pair of switches. This happens because beyond that point, increasing the length of monitoring flows causes unwanted loop in the route, therefore, those flows will not be considered.
Finally, fig.~\ref{fig:eval_barchart_Average_Percent_of_Rules_to_Switch_Rule_Capacity_Ratio_max_len_of_route_11} represent the network overhead of the proposed monitoring solution. In this plot, the average number of monitoring rules per switch versus the number of monitoring nodes are represented using left y-axis while the right y-axis shows the percentage of monitoring rules over the total capacity of the switches. Increasing the number of monitoring flows increases the network overhead, however, the percentage of the monitoring rules over the total capacity of the switch in the worse case is $0.1\%$ for Abilene and $0.3\%$ for RedIRIS which is negligible.

\section*{Acknowledgment}
This work has received funding from the EPSRC project EARL. Moreover, the work has received funding from the Cisco University Research Program Fund.

\section{Conclusion and Future Works}
In this paper, we proposed an active network monitoring architecture to measure overall link delay including both queuing and propagation delays. In this way, we inject several monitoring-flows into the network and by measuring the end-to-end delay of these flows, link delay information is inferred. To tackle the complexity, the problem is broken down into two sub-problems. Thereafter, we mathematically formulated the sub-problems and proposed two light-weight heuristic algorithms to efficiently solve the proposed problems. Simulation results support that the solution is accurate and can be used for real-world scenarios. In the proposed solution, it is considered that the monitoring nodes are predefined. Future works would be dedicated to propose a solution on monitoring nodes placement. Another field of interest is extending the solution to support Segment Routing technology.

\bibliographystyle{IEEEtran}
\bibliography{references}
\end{document}